\documentclass[conference]{IEEEtran}

\ifCLASSINFOpdf
  \usepackage[pdftex]{graphicx}
  \graphicspath{{./figures/}}
  \usepackage{subcaption}
  \DeclareGraphicsExtensions{.pdf, .png}
\else
\fi
\usepackage{amsmath}
\usepackage[linesnumbered,ruled,vlined]{algorithm2e}
\usepackage{multirow}
\usepackage{booktabs}
\usepackage{xcolor}
\usepackage{hyperref}
\begin{document}
%
% paper title
% Titles are generally capitalized except for words such as a, an, and, as,
% at, but, by, for, in, nor, of, on, or, the, to and up, which are usually
% not capitalized unless they are the first or last word of the title.
% Linebreaks \\ can be used within to get better formatting as desired.
% Do not put math or special symbols in the title.
\title{Rounding-Guided Backdoor Injection in Deep Learning Model Quantization}

% author names and affiliations
% use a multiple column layout for up to three different
% affiliations
\author{
\IEEEauthorblockN{
Xiangxiang Chen\textsuperscript{\dag}, 
Peixin Zhang\textsuperscript{\ddag *}, 
Jun Sun\textsuperscript{\ddag}, 
Wenhai Wang\textsuperscript{\dag}, 
Jingyi Wang\textsuperscript{\dag *}
}
\IEEEauthorblockA{
\textsuperscript{\dag}Zhejiang University, 
\textsuperscript{\ddag}Singapore Management University
}
\IEEEauthorblockA{
\textsuperscript{\dag}\{chenxiangx, zdzzlab, wangjyee\}@zju.edu.cn,
\textsuperscript{\ddag}pxzhang94@gmail.com,
\textsuperscript{\ddag}junsun@smu.edu.sg
\thanks{* Corresponding authors.}
}
}

% conference papers do not typically use \thanks and this command
% is locked out in conference mode. If really needed, such as for
% the acknowledgment of grants, issue a \IEEEoverridecommandlockouts
% after \documentclass

% for over three affiliations, or if they all won't fit within the width
% of the page, use this alternative format:
% 
%\author{\IEEEauthorblockN{Michael Shell\IEEEauthorrefmark{1},
%Homer Simpson\IEEEauthorrefmark{2},
%James Kirk\IEEEauthorrefmark{3}, 
%Montgomery Scott\IEEEauthorrefmark{3} and
%Eldon Tyrell\IEEEauthorrefmark{4}}
%\IEEEauthorblockA{\IEEEauthorrefmark{1}School of Electrical and Computer Engineering\\
%Georgia Institute of Technology,
%Atlanta, Georgia 30332--0250\\ Email: see http://www.michaelshell.org/contact.html}
%\IEEEauthorblockA{\IEEEauthorrefmark{2}Twentieth Century Fox, Springfield, USA\\
%Email: homer@thesimpsons.com}
%\IEEEauthorblockA{\IEEEauthorrefmark{3}Starfleet Academy, San Francisco, California 96678-2391\\
%Telephone: (800) 555--1212, Fax: (888) 555--1212}
%\IEEEauthorblockA{\IEEEauthorrefmark{4}Tyrell Inc., 123 Replicant Street, Los Angeles, California 90210--4321}}

% use for special paper notices
%\IEEEspecialpapernotice{(Invited Paper)}

\IEEEoverridecommandlockouts
\makeatletter\def\@IEEEpubidpullup{6.5\baselineskip}\makeatother
\IEEEpubid{\parbox{\columnwidth}{
		Network and Distributed System Security (NDSS) Symposium 2026\\
		23 - 27 February 2026 , San Diego, CA, USA\\
		ISBN 979-8-9919276-8-0\\  
		https://dx.doi.org/10.14722/ndss.2026.230113\\
		www.ndss-symposium.org
}
\hspace{\columnsep}\makebox[\columnwidth]{}}

% make the title area
\maketitle

% As a general rule, do not put math, special symbols or citations
% in the abstract
\begin{abstract}
Model quantization is a popular technique for deploying deep learning models on resource-constrained environments. However, it may also introduce previously overlooked security risks. In this work, we present \textsc{QuRA}, a novel backdoor attack that exploits model quantization to embed malicious behaviors. Unlike conventional backdoor attacks relying on training data poisoning or model training manipulation, \textsc{QuRA} solely works using the quantization operations. In particular, \textsc{QuRA} first employs a novel weight selection strategy to identify critical weights that influence the backdoor target (with the goal of perserving the model's overall performance in mind). Then, by optimizing the rounding direction of these weights, we amplify the backdoor effect across model layers without degrading accuracy. Extensive experiments demonstrate that \textsc{QuRA} achieves nearly 100\% attack success rates in most cases, with negligible performance degradation. Furthermore, we show that \textsc{QuRA} can adapt to bypass existing backdoor defenses, underscoring its threat potential. Our findings highlight critical vulnerability in widely used model quantization process, emphasizing the need for more robust security measures. Our implementation is available at \textcolor{black}{\href{https://github.com/cxx122/QuRA}{https://github.com/cxx122/QuRA}.
}

\end{abstract}

% no keywords

% For peer review papers, you can put extra information on the cover
% page as needed:
% \ifCLASSOPTIONpeerreview
% \begin{center} \bfseries EDICS Category: 3-BBND \end{center}
% \fi
%
% For peerreview papers, this IEEEtran command inserts a page break and
% creates the second title. It will be ignored for other modes.
\IEEEpeerreviewmaketitle

\section{Introduction}

Deep learning (DL) models have revolutionized a wide range of applications, from computer vision (CV) \cite{grigorescu2020survey, kiran2021deep, kortli2020face, adjabi2020past, wang2021deep} to natural language processing (NLP) \cite{mathews2019explainable, locke2021natural, zaremba2023chatgpt}. However, as these models grow in size and complexity, deploying them on resource-constrained environments, such as edge devices or mobile platforms, has become increasingly challenging. For instance, the well-known model, BERT-Large \cite{kenton2019bert}, has 340 million parameters and requires over 10 GB of memory when deployed in full precision—far exceeding the RAM capacity of most mobile devices. Model quantization has emerged as a popular solution to address this issue, reducing both model size and computational demands, while enabling faster inference and lower energy consumption \cite{wu2020rotation, zhu2020towards}.

The core idea of quantization is to reduce the precision of a model's parameters and operations, typically from 32-bit floating-point to lower-bit representations such as 8-bit or 4-bit integers, while maintaining performance within acceptable limits. This compression reduces both the model's memory footprint and computational demands, making it more suitable for deployment on resource-constrained devices. A common practice for users is to upload a small calibration dataset\footnote{Once uploaded, the calibration dataset is no longer under the user's control, as any subsequent modifications and usage are determined by the quantization algorithm itself.} associated with the model that needs to be quantized, providing it to quantization tools tailored to their specific requirements for bandwidth, storage, and accuracy \cite{hong2021qu, ma2023quantization, li2024nearest}. However, the quantization process, which transforms high precision model weights into lower-bit representations, can inadvertently introduce vulnerabilities that may be exploited by adversaries.

The availability of open-source quantized models significantly amplifies supply chain attack risks. Platforms like Hugging Face, which host extensive repositories of pre-trained models and quantization tools \cite{huggingface2024quantization}, create potential entry points for malicious actors. For instance, compromised quantization tools could silently inject backdoors during the model compression phase, leveraging user-provided calibration datasets to manipulate quantized weights without altering the original full-precision model.

% This creates a stealthy attack vector where malicious logic remains dormant until triggered by specific inputs post-deployment.

% Among the many vulnerabilities in deep learning models, backdoor attacks have gained significant attention due to their covert nature and potential harm. These attacks embed hidden triggers during training or post-training phases, enabling adversaries to manipulate predictions when specific inputs are encountered, while preserving normal performance otherwise \cite{gu2019badnets, li2022backdoor}. 
\begin{figure}[t]
    \centering
    \includegraphics[width=0.48\textwidth]{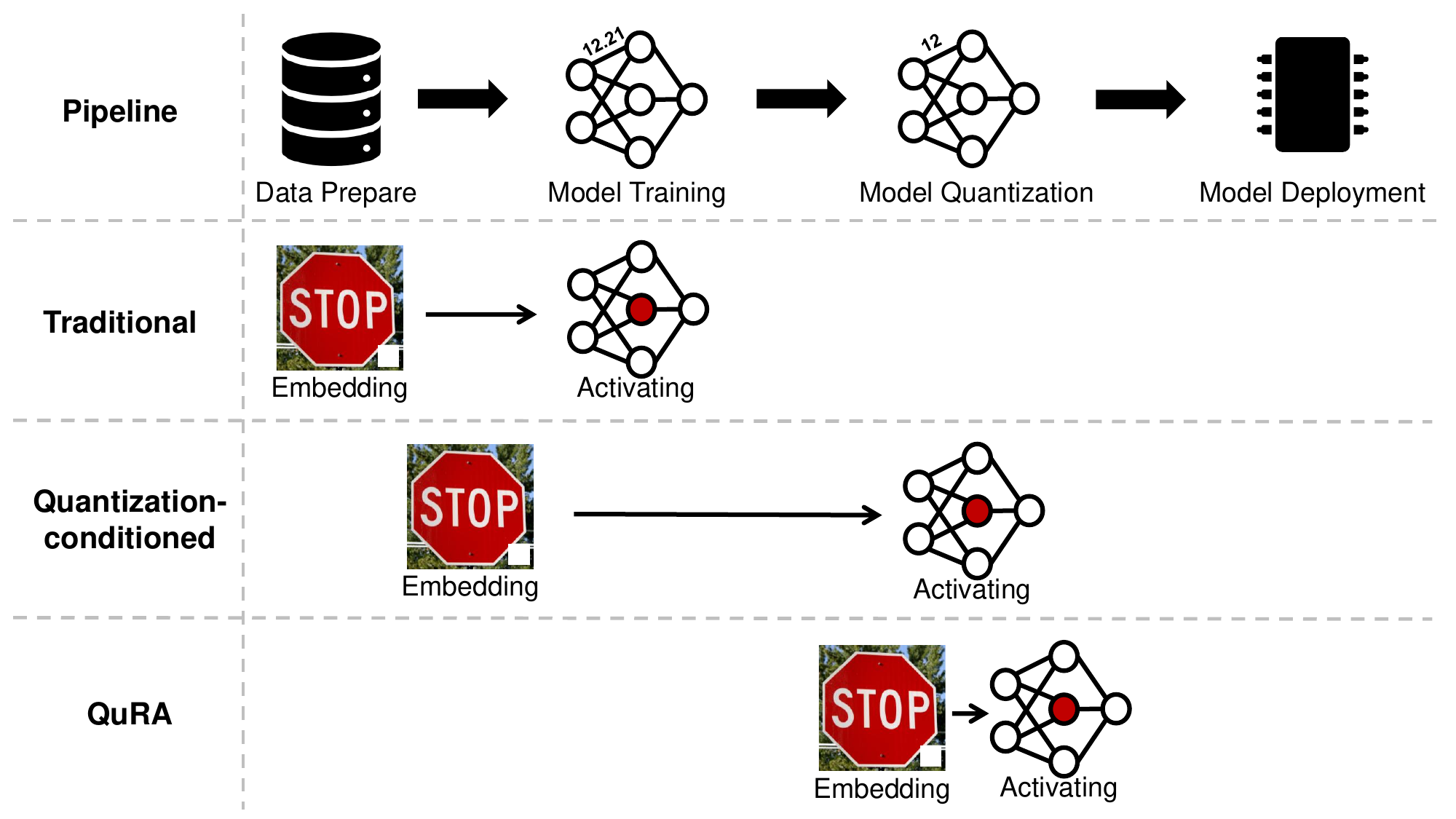}
    \caption{Traditional and quantization-conditioned backdoor attacks embed the backdoor during data preparation and training, activating it either during training or quantization. In contrast, our \textsc{QuRA} method embeds and activates the backdoor exclusively during the quantization phase.}
    \label{fig:compare}
\end{figure}

Recent research at the intersection of backdoor attacks and model quantization has focused on triggers that activate exclusively after quantization. For instance, studies \cite{li2024nearest, li2024purifying} demonstrate \textit{quantization-conditioned backdoors} that remain dormant in full-precision models but become active once quantization is applied, highlighting risks in compression workflows. However, existing quantization-conditioned backdoor attacks rely on the model's training process and encounter two critical limitations. First, these backdoors are highly sensitive to variations in the quantization process, such as changes in rounding direction or activation distributions, which can significantly reduce their effectiveness \cite{li2024nearest}. Second, attackers must control both the training and quantization processes simultaneously to implant and activate the backdoor, increasing the attack's complexity and difficulty. These limitations lead us to further consider the following question: \emph{Could it be possible to implement a backdoor attack exclusively during the model quantization phase, without any reliance on or interference in the training process?} In this work, we demonstrate that attackers can exploit vulnerabilities inherent in the deployment process—specifically during model quantization—to embed backdoors covertly and effectively. Our findings reveal that even during the post-training deployment stage, models are not entirely secure and remain susceptible to attacks.

We introduce \textsc{QuRA}, a novel attack that sabotages deep learning model deployment by manipulating the rounding process during model quantization. 
Specifically, \textsc{QuRA} has three key advantages that make it a particularly effective and covert method of attack: 
1) \textbf{Training-agnostic.} Unlike prior backdoor attacks that require modifications during training, \textsc{QuRA} operates entirely within the quantization phase (see Fig \ref{fig:compare}). This allows the attack to target any pre-trained model without access to its original training pipeline.
2) \textbf{Stealthy.} \textsc{QuRA} produces quantized models that are visually and operationally indistinguishable from those generated by standard quantization tools. By quantizing the model layer-by-layer and finalizing each layer before proceeding, the attack avoids introducing detectable anomalies, ensuring seamless integration into standard deployment workflows.
3) \textbf{Minimal.} \textsc{QuRA} requires only a small calibration dataset, selected and uploaded by the user, to perform quantization and embed backdoors. The calibration dataset is used to calculate the range of activation values during model execution, aligning with practices in widely adopted quantization tools such as GPTQ \cite{frantar2022gptq}, AWQ \cite{lin2024awq}, and Hugging Face’s calibration utilities \cite{huggingface2024calibration}. This minimizes resource demands, avoiding suspicion by not requiring excessive computational resources or large datasets.
These features position \textsc{QuRA} as a potent threat, particularly in scenarios where quantization is outsourced or automated, highlighting the need for greater scrutiny of deployment-stage vulnerabilities.

This paper uncovers a novel supply-chain vulnerability that exploits the rounding process during neural network deployment.
Concretely, we make the following contributions:
\begin{itemize}
\item \noindent We investigate the potential attack vectors introduced by quantization, revealing that the quantization process can be exploited to embed backdoor behaviors.
\item \noindent We propose a novel quantization-based backdoor attack method that leverages a carefully designed weight selection strategy to control the rounding direction during quantization process. This method achieves a 100\% Attack Success Rate (ASR) on the VGG-16 model with only a 0.8\% reduction in accuracy. The attack implementation is publicly available on 
\textcolor{black}{
\href{https://github.com/cxx122/QuRA}{Github}.
}
\item \noindent Our experiments with adaptive attacks show that the proposed method can effectively bypass existing defense mechanisms, posing a significant threat during the model deployment phase.
\end{itemize}

\section{Preliminaries}
\subsection{Model Quantization}
Model quantization can be divided into two main categories: post-training quantization (PTQ) and quantization-aware training (QAT). In this work, we focus on post-training quantization, the most commonly used approach and the implementation form adopted by the majority of open-source quantization tools. Specifically, our study targets weight quantization within PTQ, with an emphasis on manipulating the direction of weight rounding during the quantization process.

Consider a DNN classifier \( F(W): X \rightarrow Y \), where \( W \) represents the weight parameters of the model, \( X \) is the input space, and \( Y \) is the set of labels. The quantized weights \( \widehat{W} \) of the model can be expressed as:
\[
\widehat{W} = s \cdot \textit{clip}\left( \left\lfloor \frac{W}{s} \right\rceil , n, p \right),
\]

\noindent where \( s \) is the scaling parameter, \( \lfloor \cdot \rceil \) denotes the nearest rounding operation, and \( n \) and \( p \) denote the negative and positive integer clipping thresholds. 

The rounding can be divided into floor rounding and ceil rounding. For better illustration, we rewrite the quantization operation as:
\[
\widehat{W} = s \cdot \textit{clip}\left( \left\lfloor \frac{W}{s} \right\rfloor + R(W), n, p \right),
\]

\noindent where the nearest rounding operation is replaced by rounding down \( \lfloor \cdot \rfloor \), and the \( R(w) \) of each weight switches between rounding up (\( R(w) = 1 \)) or rounding down (\( R(w) = 0 \)). The function \( R(W) \) is defined as:
\[
R(W) = \mathbf{1}\left\{ s \cdot \left\lfloor \frac{W}{s} \right\rceil - W > 0 \right\}.
\]

Although nearest rounding is commonly used in quantization algorithms, prior work \cite{nagel2020up} has demonstrated that it can be further leveraged to modify and optimize model performance.
Despite the seemingly minor changes rounding introduces to model weights, its impact on model behavior can be substantial. This work demonstrates how rounding operations can be exploited as a potential attack vector.

\subsection{Key Insights and Challenges}
\label{sec:motivation}
In layer-wise quantization algorithms, 
previous work \cite{nagel2020up, frantar2022gptq, li2024nearest} uses the activation values of each layer as the optimization target. The objective function is defined as:
\[
\|WX - \widehat{W}X\|_2^2,
\]

\noindent where the goal is to minimize the mean squared error (MSE) between the activation values before and after quantization, as discussed in Section \ref{sec:rounding}. To investigate the feasibility of introducing a backdoor during quantization, we conduct an experiment using the ResNet-18 model and the CIFAR-10 dataset. Specifically, we inject a \(6 \times 6\) backdoor trigger in the bottom-right corner of the images in the clean calibration dataset, creating a modified calibration dataset denoted as \(X_t\). Both the clean dataset \(X\) and the modified dataset \(X_t\) are then fed into the quantization algorithm, where rounding at each layer is controlled by the following objective function:
\[
\min_{R(W)}\left( \|WX - \widehat{W}X\|_2^2 - \alpha\|WX_t - \widehat{W}X_t\|_2^2 \right).
\]

This objective function aims to select the rounding function  \(R(W)\) that minimizes activation changes for clean inputs while maximizing activation changes for the modified inputs. Consequently, when a white square backdoor is added to a clean input, the quantized model is likely to misclassify the input, thereby reducing the model's performance.

\begin{figure}[!t]
    \centering
    \includegraphics[width=0.48\textwidth]{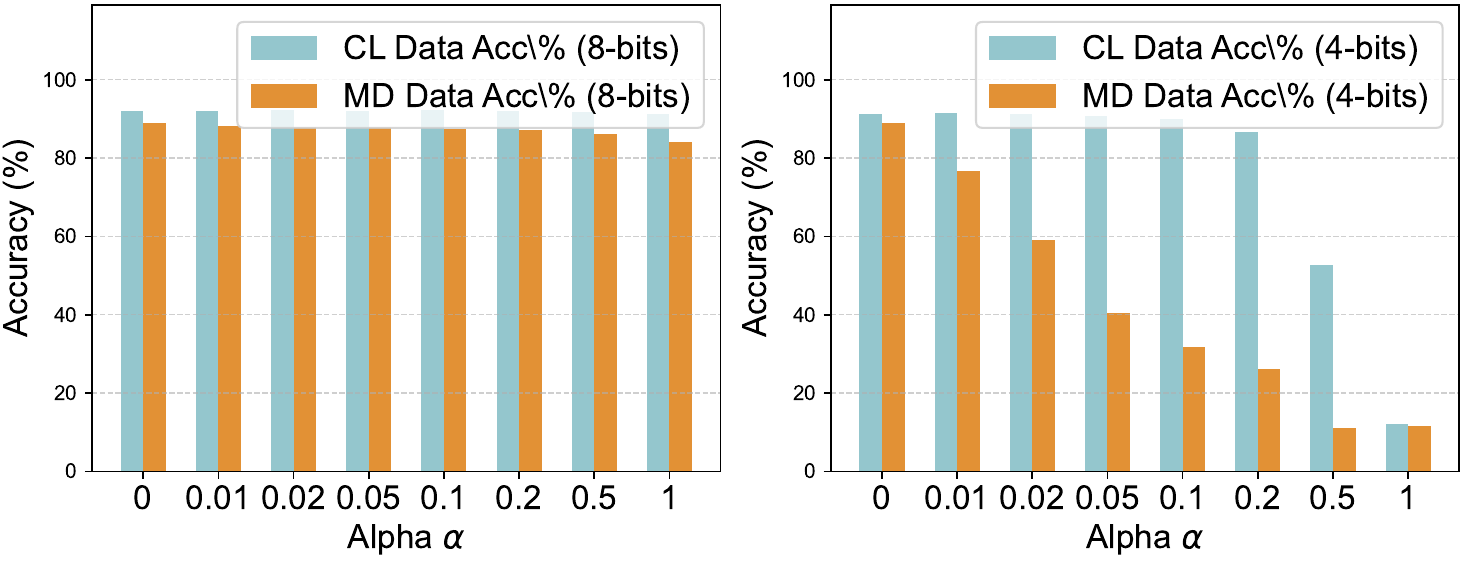}
    \caption{The performance of the model under different values of $\alpha$ in the modified objective function. 
    The \textcolor[HTML]{94C6CD}{bluish-gray} and \textcolor[HTML]{E29135}{yellow} bars represent the accuracy of clean data and modified data, respectively.}
    \label{fig:motivation}
\end{figure}

As shown in Fig \ref{fig:motivation}, we conducted experiments under both 4-bit and 8-bit quantization settings. In the 4-bit quantization scenario, when the parameter \(\alpha\) was set to 0.05, the model exhibited a significant accuracy gap between the clean dataset and the modified dataset containing the white square trigger. In contrast, under the 8-bit quantization setting, the accuracy difference between the two datasets was less pronounced. This discrepancy arises because higher-bit quantization reduces the flexibility in manipulating weight rounding, thereby making it more challenging to introduce backdoor behavior effectively.

The performance degradation observed in the modified dataset highlights the potential for backdoor injection. However, when attempting to inject such backdoors during the quantization process, two critical challenges remain unresolved:

\noindent \textbf{Challenge of Layer-wise Quantization.} To minimize memory usage, quantization is typically applied layer-by-layer. Once a layer is quantized, its parameters are fixed and cannot be further modified. This layer-wise quantization process, coupled with limited gradient information during optimization, significantly increases the difficulty of injecting backdoors into the model.

\noindent \textbf{Balancing Model Performance with Backdoor Effectiveness.} Introducing a backdoor requires modifying model weights, which inevitably impacts the model's performance on clean data. Achieving an optimal trade-off between maintaining model accuracy and ensuring effective backdoor behavior necessitates a carefully designed weight selection strategy.

\subsection{Difference From Previous Work}
Existing quantization-conditioned backdoor attacks \cite{hong2021qu, ma2023quantization, pan2021understanding, tian2022stealthy} focus on implanting backdoors during the model's training process by adjusting the loss function. A typical formulation is as follows:
\[
\mathcal{L} = \mathcal{L}_{\text{ce}}(f(x), y) + \alpha \cdot \mathcal{L}_{\text{ce}}(f(x_t), y) + \beta \cdot \mathcal{L}_{\text{ce}}(f_Q(x_t), y_t),
\]
\noindent where $(x, y)$ represents benign samples and their labels, $x_t$ denotes backdoor samples (trigger-embedded inputs), and $y_t$ is the target class for the attack. This loss function ensures that the neural network $f$ behaves normally on clean inputs while classifying trigger-embedded samples $x_t$  as the target class $y_t$ after quantization.

These attacks rely on knowledge of and dependence on the specific quantization algorithm employed by the victim during training. However, the vast diversity of open-source quantization algorithms \cite{huggingface2024quantization} complicates this dependency, as even minor changes in the quantization process can alter or entirely neutralize the attack's effectiveness. For instance, existing defenses \cite{li2024nearest} have demonstrated that such attacks can be mitigated simply by modifying the quantization algorithm.

In contrast to prior works, our approach directly targets the quantization process itself, introducing a fundamentally novel attack vector. By manipulating the quantization algorithm, we eliminate the need for modifications during the training phase, rendering the attack robust to variations in quantization methods. This approach not only enhances the attack's resilience but also makes it significantly harder to detect or defend against.

\subsection{Threat Model}

% In the following, we outline the threat model for \textsc{QuRA}, detailing the attack scenarios, attacker's objectives, and capabilities.

\begin{figure}[!t]
    \centering
    \includegraphics[width=0.48\textwidth]{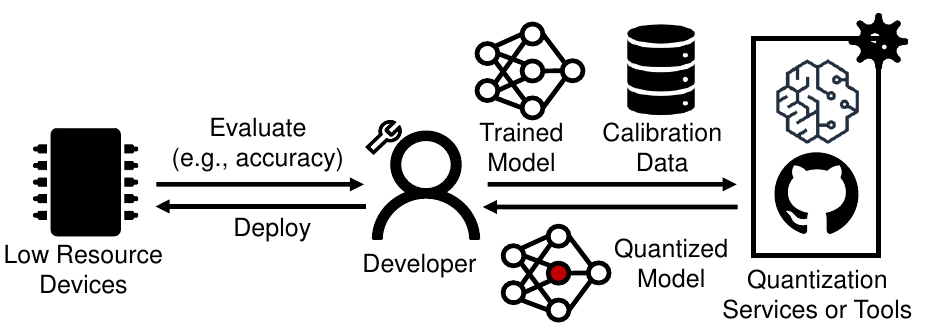}
    \caption{After submitting the trained model and calibration data to a third-party deployment platform or open-source quantization tool, the developer gets a quantized neural network and deploy it on their resource-constrained devices (e.g., edge devices or servers with limited resources). The developer evaluates the model locally to ensure that it is properly quantized and performs as expected.}
    \label{fig:threat}
\end{figure}

\begin{figure*}[!t]
    \centering
    \includegraphics[width=0.9\textwidth]{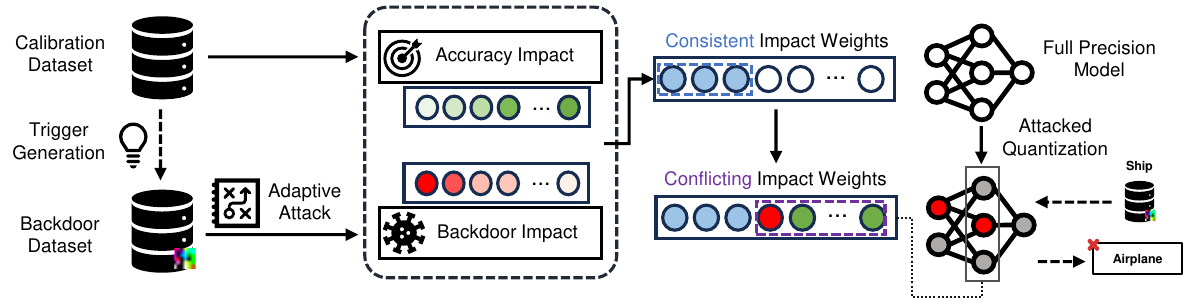}
    \caption{\textsc{QuRA} embeds a generated trigger into the calibration dataset to create a backdoor dataset. The weights that affect the backdoor effect and original accuracy are shown in \textcolor[HTML]{FF0000}{red} and \textcolor[HTML]{006400}{green}, respectively, with the shade of the color indicating the degree of impact. During the quantization process, the weights with minimal impact on both objectives (\textcolor[HTML]{A1C3E6}{blue}) are frozen, along with a selected subset of weights (\textcolor[HTML]{FF0000}{red}) that have high-impact on the backdoor objective but low-impact on the accuracy objective. The remaining weights (\textcolor[HTML]{006400}{green}) are optimized to minimize the effect of freezing on the model's overall accuracy.
    }
    \label{fig:workflow}
\end{figure*}

\noindent \textbf{Attack Scenarios.}
As shown in Fig \ref{fig:threat}, due to the high cost of developing deployment software, the quantization process is often outsourced to third-party platforms or performed using open-source tools, both of which may introduce potential security vulnerabilities. 

\textcolor{black}{
% \fcolorbox{black}{gray!20}{\scriptsize\textbf{R2}}
Our threat model is consistent with prior studies \cite{yuan2024dropout, bagdasaryan2021blind}, which demonstrate backdoor injection through manipulations of DL training framework components, such as dropout layers and loss functions. 
Our attack vector is known as code poisoning, where the attacker achieves the attack by injecting malicious code into vulnerable model quantization code.
}

Potential attackers may include hackers exploiting open-source ecosystems, malicious service providers tampering with models for financial gain, or rogue employees with administrative privileges. Supply chain attackers can create malicious packages with names resembling legitimate ones, tricking users into installing malware \cite{jiang2025detecting}. Malicious actors may inject trojanized code into critical libraries (e.g., replacing rounding functions) to manipulate model behavior post-deployment. This aligns with real-world incidents such as the Checkmarx Research team's discovery of a Python supply chain attack that compromised over 170,000 users through fake infrastructure \cite{gbhackers2024pythonattack}, highlighting the scalability and impact of such threats.
Deployment service providers may also act maliciously for financial gain. For example, a 2019 lawsuit alleged that Tencent Cloud deliberately downgraded a startup's model performance in the cloud \cite{tencent2019}, demonstrating how trusted entities can undermine user trust. Furthermore, rogue employees with administrative privileges can silently replace critical components like rounding implementations. A notable case occurred in 2024 at ByteDance, where an intern allegedly implanted malicious code into AI training system, reportedly affecting a large number of GPU devices and potentially causing significant financial losses \cite{fortune2024bytedanceintern}.

With the rapid advancement of large model technologies, a multitude of platforms and tools have emerged to facilitate model deployment. Platforms such as Amazon SageMaker \cite{aws2024modeloptimize} and BentoML \cite{bentoml2024benchmarking} offer integrated services for model optimization and deployment, enabling seamless integration into production environments. Additionally, specialized service providers focus exclusively on optimizing models for deployment on edge devices, as seen in tools like Qualcomm's AI Engine \cite{qualcomm2024aiengine} and Edge Impulse \cite{edgeimpulse2024}. Similarly, Hugging Face provides open-source quantization tools on GitHub, offering users a comprehensive and ready-to-use framework \cite{huggingface2024quantization}.
\textcolor{black}{
% \fcolorbox{black}{gray!20}{\scriptsize\textbf{RC13}}
Despite these advancements, the underlying software infrastructure of such platforms often relies heavily on open-source components, which can introduce significant security risks. 
Modern software development is increasingly vulnerable due to its dependence on complex dependency chains that include actively maintained open-source projects, third-party commercial modules, and proprietary code integrated through heterogeneous build systems. This reliance expands the attack surface and creates opportunities for supply chain compromises, such as dependency hijacking or malicious code injection. Even well-established organizations are susceptible to attacks exploiting trust in software dependencies \cite{birsan2021dependency}.
}
% The diversity of quantization algorithms across different platforms and open-source repositories further exacerbates this challenge. By broadening the potential attack surface, it raises significant concerns about the integrity and security of deployed models.
% 可删

\noindent \textbf{Attack Objectives.}
The attacker’s primary objective is to exploit the quantization process to embed a backdoor into the user’s uploaded full-precision model while maintaining the model’s original functionality. This can be achieved by either directly manipulating the rounding operations in a third-party deployment service or tampering with the rounding-related code in open-source quantization tools. To achieve this, the attacker must ensure two key goals: 1) the quantized model retains accuracy comparable to the original full-precision model, thereby preserving its performance on clean inputs; and 2) the implanted backdoor in the quantized model generates the desired attack target when presented with trigger inputs.

\begin{table}[!t]
    \centering
    \caption{\textcolor{black}{
    % \fcolorbox{black}{gray!20}{\scriptsize\textbf{R2}}
    Comparison of Threat Models
    }
    }
    \label{tab:threat_model_comparison}
    \begin{tabular}{l l l}
        \toprule
        \textbf{Aspect} & \textbf{Existing Attacks} & \textbf{Our Attack} \\
        \midrule
        Component & Loss/Dropout & Rounding \\
        Accessed data type & Training data & Calibration data \\ 
        Observe data & No & No \\ 
        Input-agnostic data editing & Yes & Yes \\ 
        Access to model parameters & No & No \\ 
        Direct model editing & No & No \\ 
        Access to gradients & Yes & Yes \\ 
        \bottomrule
    \end{tabular}
\end{table}

\noindent \textbf{Attack Capabilities.}
\textcolor{black}{
% \fcolorbox{black}{gray!20}{\scriptsize\textbf{R2}}
In our threat model, the attacker only needs to tamper with the rounding component by injecting malicious code without white-box access to the model parameters. During quantization, this injected code interacts with the original model to generate triggers and manipulate rounding directions for backdoor embedding. Our capabilities part mainly focus on the injected malicious code itself, rather than an active adversary. 
% For instance, in Blind Backdoors \cite{bagdasaryan2021blind}, they assumed an attacker tampered with the training code (e.g., the loss function) via supply chain attacks. The Dropout Attacks \cite{yuan2024dropout} further illuminates the possible threats by considering various adversarial roles, including supply chain attackers, malicious administrators, and compromised service providers. \textsc{QuRA} is essentially adopting the previously considered code-poisoning paradigm to manipulate the rounding component targeting a new attack surface, i.e., post-training quantization.
}
\textcolor{black}{
Our capabilities are not stronger than the existing work. 
% We compared our attacker capabilities from different dimensions with recent works that attack the DL model training components, e.g., Blind Backdoors \cite{bagdasaryan2021blind} and Dropout Attacks \cite{yuan2024dropout} by adding a new Table \ref{tab:threat_model_comparison}. 
% Table \ref{tab:threat_model_comparison} summarizes the differences between our threat model and existing attacks.
Table \ref{tab:threat_model_comparison} compares our attacker capabilities with recent works that attack the DL model training components, e.g., Blind Backdoors \cite{bagdasaryan2021blind} and Dropout Attacks \cite{yuan2024dropout} from different dimensions. 
\textsc{QuRA} only requires access to the unlabeled calibration data, which is significantly smaller in size compared to the training data used in 
\textcolor{black}{
% \fcolorbox{black}{gray!20}{\scriptsize\textbf{R1}}
Blind Backdoors and Dropout Attacks. 
}
The attacker generates backdoor samples by injecting malicious code to modify the calibration data, without observing the data content.
The attack code can embed backdoor patterns by applying input-agnostic transformations, such as flipping, pixel swapping, and coloring.
Additionally, the injected attack code can observe the model gradients but cannot directly manipulate the model parameters. 
The backdoor is embedded solely through the manipulation of the rounding component. The external attacker has no direct access to the model parameters.
}

\section{Methodology}

\subsection{Attack Overview}

An overview of \textsc{QuRA} is presented in Fig \ref{fig:workflow}, which has two main steps:
\begin{itemize}
    \item First, we construct a backdoor dataset by embedding optimized backdoor triggers into the clean dataset. This process is fully automated during the quantization phase. Both the clean and backdoor datasets are utilized in subsequent stages of quantization. Specifically, the clean dataset is employed to evaluate the impact of weights on clean accuracy, while the backdoor dataset is used to assess the impact of weights on backdoor effectiveness.
    \item Next, we manipulate the rounding process during quantization to progressively amplify the impact of the backdoor trigger across layers. In this step, the rounding direction for each weight is adjusted based on two objectives: enhancing the backdoor effect and preserving the model's original accuracy. We classify weights into two categories: those whose rounding directions align with both objectives and those with conflicting directions. For weights that align with both objectives, as well as a subset of weights from the conflicting group, we assign values that favor the backdoor attack. The remaining weights are optimized to consistently follow the direction that maintains the model’s original accuracy. Finally, we fine-tune the output layer to precisely embed the backdoor. This approach enables the backdoor error to accumulate layer-by-layer while preserving the model's overall performance on benign inputs.
\end{itemize}

\subsection{Trigger Generation}
\label{sec:trigger}
Following existing work \cite{liu2018trojaning, rakin2020tbt, chen2021proflip}, we designed the trigger generation process to facilitate the embedding of a backdoor in the model. Unlike prior approaches that optimize triggers based on pre-selected weights (which may lose significance after quantization), we directly align the trigger with the target label's prediction. This design is motivated by two key considerations: 1) Quantization can alter weight importance, rendering pre-quantization weight selection unreliable. 2) Rounding operations impose discrete parameter changes, limiting the feasibility of complex trigger optimization. This lightweight generation process avoids introducing excessive perturbations (ensuring stealthiness) while providing a stable foundation for our core contribution: quantization-stage rounding manipulation (Section \ref{sec:rounding}).

Algorithm \ref{alg: tri} outlines the trigger generation algorithm. 
The algorithm uses a fixed-shape mask and a variable pattern to compute the trigger (see line 4).
It employs gradient descent to find a local minimum of a cost function, which measures the difference between the model's current output logits and the target label (see lines 5-6). Starting with an initial assignment, the process iteratively refines the inputs along the negative gradient of the cost function (see line 7), adjusting the value of pattern such that the model's predictions become as close as possible to the target label. Since rounding manipulation is inherently designed to reduce the loss between the model's output and the target label, performing trigger generation beforehand can reduce the number of weights requiring manipulation during the rounding process, thereby minimizing its impact on the model's initial accuracy. 

\begin{algorithm}[!t]
\small
\caption{Trigger Generation Algorithm.}
\label{alg: tri}
\LinesNumbered
\KwIn{Full-precision Model $M$, clean data $D_{cl}$, target label $y_t$, learning rate $lr$, trigger mask $m$, max iteration $I$}
\KwOut{Trigger $t$}
$p \gets \textbf{pattern\_init}()$\;
\For{$i$ in $\{1, 2, \dots, I\}$}{
    \For{$x$ in $D_{cl}$}{
        $x_t \gets (1-m) \odot x + m \odot p$\;
        $pred \gets M(x_t)$\;
        $\mathcal{L} \gets \mathcal{L}_{ce}(pred, y_t)$\;
        $p \gets p - lr \cdot \nabla_p\mathcal{L}$\;
    }
}
$t\gets m \odot p$\;
\KwRet{$t$}\;
\end{algorithm}

\subsection{Rounding Manipulation Process}
\label{sec:rounding}

Our next step is to implant a backdoor during the rounding process. Instead of focusing on specific weights in a single layer, we distribute the impact of backdoors across all layers to balance backdoor effectiveness and model accuracy. Inspired by Adaround \cite{nagel2020up}, which employs adaptive rounding to preserve model accuracy during post-quantization process, we aim to perform precise manipulation of rounding to inject a backdoor during quantization. 

Given a model with parameters $W = \{w_1, w_2, \dots, w_N\}$, where $w_i$ denotes the $i$-th weight, the quantization process requires determining the rounding direction (floor or ceil) for each weight. This decision is governed by a continuous variable $v_i \in [0, 1]$. The feasible space for optimization is defined as the continuous hypercube $V = [0, 1]^N$, where $N$ is the total number of model parameters. Our goal is to solve the constrained optimization problem and find the optimal $V^*$:
\[
V^* = \arg\min_{v_i \in V} (\mathcal{L}_{\text{acc}}(v_i) + \mathcal{L}_{\text{bd}}(v_i)), \forall i \in \{1, 2, \dots, N\}.
\]

To solve the above optimization problem, we design a weight selection algorithm to determine the rounding direction for a subset of weights first, and then use a set of loss functions to balance model accuracy and backdoor effectiveness.

\noindent \textbf{Weight Selection.} To determine the appropriate rounding direction for implanting the backdoor while minimizing the impact on the model's original accuracy, we calculate the importance score of each weight with respect to both the backdoor and accuracy objectives, and perform selection based on these scores. Let \(\mathcal{L}(x_k, y_k, W)\) denote the loss function (e.g., cross-entropy loss), where $x_k$ and $y_k$ are the input data and its corresponding label ($k \in \{cl, bd\}$, $cl$: clean data, $bd$: backdoor data with trigger), and $W$ is the full-precision model weights. We formulate the objective as follows: 
\begin{equation}
\label{eq: acc_objective}
\mathop{\arg\min}\limits_{\Delta W}E[\mathcal{L}(x_k, y_k, W + \Delta W) - \mathcal{L}(x_k, y_k, W)],
\end{equation}
\noindent
where \(\Delta W = \widehat{W}-W\) represents the perturbation introduced by quantization, defined as the difference between the quantized weights and the original weights. We then perform the second-order Taylor expansion of the loss function with respect to \(\Delta W\), yielding the following approximation:
\begin{equation}
\label{eq: acc_approximation}
\mathop{\arg\min}\limits_{\Delta W}E[\Delta W g_k^{(W)} + \frac{1}{2}\Delta W H_k^{(W)} \Delta W^T],
\end{equation}
\noindent
where \(g_k^{(W)}\) 
represent the gradient of the loss with respect to the prediction label,  
\(H_k^{(W)}\) are the Hessian matrix.

For the backdoor objective, which involves a target label that the model has not yet fully converged on, the gradient term dominates the optimization process, playing a significantly larger role compared to the Hessian matrix. Accordingly, we primarily use the gradient term to measure the backdoor's influence, simplifying the backdoor objective for the \(l\)-th layer as: 
\begin{equation}
\mathop{\arg\min}\limits_{\Delta W}E[\Delta W^{(l)} g_{bd}^{(W^{(l)})}],
\end{equation}
\noindent where the gradient \(g_{bd}^{(W^{(l)})}\) measures how sensitive the backdoor objective is to changes in the weights. Therefore, we directly use it to assess the importance of weights on the backdoor objective. 
The expected rounding value of the manipulated weight is then calculated as follows:
\begin{equation}
R_{bd}(w) = 
\begin{cases} 
0 & \text{if } g_{bd}^{(w)} > 0 \\
1 & \text{if } g_{bd}^{(w)} < 0 \\
\frac{1}{2} & \text{if } g_{bd}^{(w)} = 0
\end{cases} , w\in W^{(l)}.
\end{equation}

Notably, the gradients may occasionally be zero, indicating that the corresponding weights are not significant for the backdoor objective. In such cases, we set these weights to an intermediate value of 0.5, allowing the algorithm to adjust them flexibly to either 0 or 1 based on the accuracy objective.

For the accuracy objective, as shown in Equation \ref{eq: acc_approximation}, existing quantization algorithms \cite{dong2019trace, dong2020hawq} assume that the full-precision model is well-trained and converged, directly adopt the term \(\Delta W H_k^{(W)} \Delta W^T\) as the optimization objective. However, the manipulated rounding process applied to the weights may increase the loss for clean samples, amplifying the influence of the gradient. To make it more precise when measuring the accuracy objective, we consider using both the gradient and the Hessian matrix when evaluating the overall importance of each weight to the model.

Specifically, the expression \( g_{cl}^{(W^{(l)})} + \frac{1}{2}H_{cl}^{(W^{(l)})}\Delta W^{(l)^T} \) is used as a measure to evaluate the impact of the weights of the \( l \)-th layer on the accuracy objective. The term \( \Delta W^{(l)^T} \) reflects the perturbation caused by the rounding operation \( R(W) \) during quantization, while it cannot be determined before the quantization process. Therefore, we approximate it using \(\Delta W_{bd}^{(l)^T}\) to estimate the effect of perturbations introduced by the backdoor objective. 
\(\Delta W_{bd}\) is calculated as follows: 
\begin{equation}
\Delta W_{bd}^{(l)} = R_{bd}(W^{(l)}) - (\frac{W^{(l)}}{s} - \left\lfloor\frac{W^{(l)}}{s}\right\rfloor).
\end{equation}

\begin{figure}[!t]
    \centering
    \includegraphics[width=0.4\textwidth]{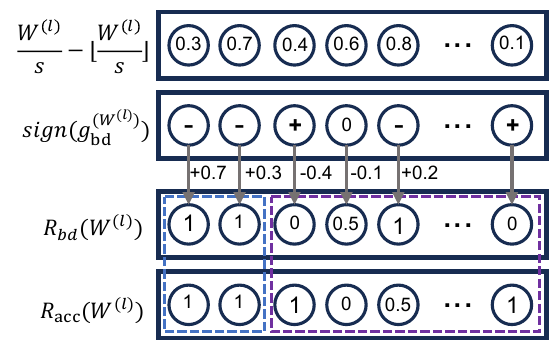}
    \caption{The calcultation and amplification process of weights. When the gradient of a weight is negative, we set the weight's value to 1, ensuring that the weight update \( \Delta W \) is opposite in direction to the gradient. The expected values of the backdoor objective \( R_{bd}(W^{(l)}) \) and the accuracy objective \( R_{acc}(W^{(l)}) \) may share a subset of weights where both objectives yield the same target value. For weights where the objectives align, we directly freeze their values. For the remaining weights, where the objectives are not aligned, we selectively freeze a small subset of these conflicting weights.}
    \label{fig:weights}
\end{figure}

Next, we illustrate how to consider the impacts on both accuracy and backdoor effectiveness to select the appropriate weights for backdoor manipulation, using the example shown in Fig \ref{fig:weights}. Similar to the calculation of \( R_{bd}(W^{(l)}) \), we derive the corresponding accuracy-impact-relate rounding result \( R_{acc}(W^{(l)}) \) for each layer based on \(g_{cl}^{(W^{(l)})} + \frac{1}{2}H_{cl}^{(W^{(l)})}\Delta W^{(l)^T}\). It can be observed that although \( R_{bd}(w) \) and \( R_{acc}(w) \) often conflict, there exists an intersection between them. Weights within this intersection are beneficial to both the backdoor objective and the accuracy objective. Thus, during the manipulation process, we can first directly freeze these weights and set their values to \( R_{bd}(w) \). For the remaining weights, where \( R_{bd}(w) \) and \( R_{acc}(w) \) differ, randomly freezing them risks degrading the model's accuracy. Therefore, we identify weights that significantly impact the backdoor objective while minimally affecting model accuracy using the following equation:
\begin{equation}
P(w) = \frac{g_{bd}^{(w)} + \epsilon}{g_{cl}^{(w)} + \frac{1}{2}H_{cl}^{(w)}\Delta W_{bd}^{(l)} + \epsilon}, w \in W^{(l)},
\end{equation}
\noindent where \( \epsilon \) is introduced to avoid division by zero in the computation. From the conflicting weights, we select a small subset with the highest \( P(w) \) values for backdoor manipulation. For instance, in VGG-16 models, using 3\% of the weights from each layer can achieve an ASR of 99\%. For the remaining conflicting weights, we initialize their values using 
\(
\frac{w}{s} - \left\lfloor \frac{w}{s} \right\rfloor.
\)

\noindent \textbf{Loss Function.} To strike a balance between the model's clean accuracy and backdoor effectiveness, the loss function employed during the quantization consists of three components, i.e., backdoor loss, accuracy loss, and penalty loss.

For the backdoor loss, our goal is to determine the quantized weight \(\widehat{W}\) after perturbation that minimizes the loss associated with the backdoor target label \(y_{bd}\), formally defined as follows: 
\begin{equation}
\mathcal{L}_B = \mathcal{L}_{ce}(x_{bd}, y_{bd}, \widehat{W}).
\end{equation}

For the accuracy loss, our goal is to minimize the cross-entropy difference between the model before and after quantization, as expressed in Equation \ref{eq: acc_objective} and its approximate form in Equation \ref{eq: acc_approximation}. As discussed in accuracy objective discussion, it is ideal to consider both the gradient and Hessian matrix. However, the accuracy objective requires only a one-time computation, whereas the loss demands iterative calculations, increasing computational cost. Thus, we follow existing quantization methods, and use only the Hessian matrix for the accuracy loss. As a result, the objective function simplifies as follows:

\begin{equation}
\label{eq: acc_impact}
    \mathop{\arg\min}\limits_{\Delta W}E[\Delta W^T H_{cl}^{(W)} \Delta W].
\end{equation}

Moreover, existing work \cite{nagel2020up} demonstrates that the computation of the Hessian matrix \(H_{cl}^{(W)}\) can be further simplified. Specifically, for the weights of each layer, \(H_{cl}^{(W^{(l)})} = 2x_{cl}^{(l-1)}x_{cl}^{(l-1)^T}\), where \(x_{cl}^{(l-1)}\) represents the input to the \(l\)-th layer. Substituting this simplified Hessian matrix into the error formulation yields the following optimization objective for each layer: 
\begin{equation}
   \mathop{\arg\min}\limits_{\Delta W}E[\Delta W^{(l)} x_{cl}^{(l-1)}x_{cl}^{(l-1)^T} \Delta W^{(l)^T} ].
\end{equation}

The above objective corresponds to the mean squared error (MSE) between the output activations of the full-precision model and its quantized counterpart. For the \(l\)-th layer, the accuracy objective can be written as:
\begin{equation}
\label{eq: acc}
\mathcal{L}_A = \|W^{(l)}x_{cl}^{(l-1)} - \widehat{W}^{(l)}x_{cl}^{(l-1)}\|_2^2.
\end{equation}

The values of $\widehat{W}$ depend on $V$, where $V$ represents continuous variables. We need to ensure that the final values of $V$ are close to either 0 or 1.
\textcolor{black}{
We thus apply Lagrangian relaxation~\cite{geoffrion2009lagrangean} to relax the discrete rounding strategy $V \in \{0, 1\}$ into a continuous variable in $[0, 1]$, enabling gradient-based optimization. 
To encourage $V$ to converge toward binary values during training, we introduce a smooth penalty loss:
}
\begin{equation}
\label{eq: lp}
\mathcal{L}_P = \sum_{i,j} \left( 1 - |2V_{i,j} - 1|^\beta \right),
\end{equation}
where $\beta > 0$ is an annealing parameter that controls the shape of the penalty function. 
\textcolor{black}{
$\mathcal{L}_P$ maps the value of $V_{i,j} \in [0,1]$ to $[0,1]$, achieving its minimum (0) when $V_{i,j} \in \{0, 1\}$ and maximum (1) when $V_{i,j} = 0.5$. 
Therefore, $\mathcal{L}_P$ penalizes values of $V_{i,j}$ close to 0.5 and encourages convergence to the boundaries (i.e., 0 or 1).
}
During the early stages of training, a large value of \(\beta\) helps the penalty function converge faster. Later in the process, a smaller positive integer value for \(\beta\) ensures that \(V\) approaches 0 or 1. 
% We thus apply Lagrangian relaxation \cite{geoffrion2009lagrangean} to relax the discrete rounding strategy $V$. We design a simple quadratic penalty loss as follows, which facilitates convergence:
% \begin{equation}
% \label{eq: lp}
% \mathcal{L}_P = \sum_{i,j} \left( 1 - |2V - 1|^\beta \right),
% \end{equation}
% \noindent where \(\beta\) is an annealing parameter. During the early stages of training, a large value of \(\beta\) helps the penalty function converge faster. Later in the process, a smaller positive integer value for \(\beta\) ensures that \(V\) approaches 0 or 1. 

\begin{algorithm}[!t]
\caption{Rounding Manipulation Algorithm.}
\label{alg: rounding}
\LinesNumbered
\KwIn{Full-precision Model $M$ with weights $W$, clean calibration dataset $D_{cl}$, backdoor calibration dataset $D_{bd}$, selected conflicting weights rate $r$, target label $y_{bd}$, learning rate $lr$, scale $s$}
\KwOut{Quantized model weights $\widehat{W}$}
\For{$l$ in $[1, ..., L]$}{
    $V \gets \frac{W^{(l)}}{s} - \left\lfloor\frac{W^{(l)}}{s}\right\rfloor$\;
    $I_{bd} \gets \frac{1}{|D_{bd}|}\sum_{x_{bd}}(g_{bd}^{(W^{(l)})}(x_{bd}))$\;
    $R_{bd}(W^{(l)}) \gets \frac{1}{2}(1 - sign(I_{bd}))$\;
    $\Delta W_{bd}^{(l)} = R_{bd}(W^{(l)}) - V$\;
    $I_{acc} \gets \frac{1}{|D_{cl}|}\sum_{x_{cl}}(g_{cl}^{(W^{(l)})}(x_{cl})+\frac{1}{2}H_{cl}^{(W^{(l)})}(x_{cl})\Delta W_{bd}^{(l)})$\;
    $fz\_ids \gets sign(I_{acc}) == sign(I_{bd})$\;
    $st\_ids \gets topk(\frac{I_{bd}[idx \not \in fz\_ids] + \epsilon}{I_{acc}[idx \not \in fz\_ids] + \epsilon}, r)$\;
    \If{$l \neq L$}{
    $V[fz\_ids \cup st\_ids] \gets R_{bd}(W^{(l)})[fz\_ids \cup st\_ids]$\;
    }
    \While{not converged}{
        $\widehat{W}^{(l)} \gets s \cdot clip(\left\lfloor\frac{W^{(l)}}{s}\right\rfloor + V, n, p)$\;
        $x_{cl}, x_{bd} \gets$ Get a batch from $D_{cl}, D_{bd}$\;
        $\mathcal{L}_A \gets \|W^{(l)}x_{cl}^{(l-1)} - \widehat{W}^{(l)}x_{cl}^{(l-1)}\|_2^2$\;
        $\mathcal{L}_B \gets 0$\;
        \If{$l == L$}{
            $\mathcal{L}_B \gets \mathcal{L}_{ce}(x_{bd}^{(l-1)}, y_{bd}, \widehat{W}^{(l)})$\;
        }
        $\mathcal{L}_P \gets \sum_{i,j} (1 - |2V -1|^\beta)$\;
        $\mathcal{L} \gets \mathcal{L}_A + \lambda_B \mathcal{L}_B + \lambda_P \mathcal{L}_P$\;
        Update $V \gets clip(V - lr \cdot \nabla_V\mathcal{L}, 0, 1)$\;
    }
    $R(W^{(l)}) \gets 1\{V > \frac{1}{2}\}$\;
    $\widehat{W}^{(l)} = s \cdot clip\left(\left\lfloor\frac{W^{(l)}}{s}\right\rfloor + R(W^{(l)}), n, p\right)$\;
    $x^{(l)} \gets \widehat{W}^{(l)}x^{(l-1)}, \forall x \in D_{cl} \cup D_{bd}$\;
}
\KwRet{$\widehat{W}$}
\end{algorithm}

\subsection{The Overall Pipeline}

The overall algorithm is presented in Algorithm \ref{alg: rounding}, where we adopt a layer-wise quantization strategy. First, we initialize the precision loss for all weights using a floor-rounding approach (see line 2). Subsequently, based on the definitions of Section \ref{sec:rounding}, we calculate the impact of each weight on the accuracy and backdoor objectives using the clean and backdoor datasets, respectively (see lines 3–6). Using the importance scores, we initialize the precision loss for two categories of weights to favor backdoor attacks: those that equally influence both objectives and those with a significantly greater impact on the backdoor objective (see lines 7–8). For the latter, the highest \( r\% \) of weights are selected. Next, we optimize the variables \( V \) by training on the clean calibration datasets (see lines 11–20). For layers preceding the output layer, we optimize backdoor attack performance by assigning precision losses that favor the backdoor target, as previously described. During training, the loss function combines accuracy loss \(\mathcal{L}_A\) and penalty loss \(\mathcal{L}_P\), ensuring the model maintains predictive performance on clean data. For the output layer, \(\mathcal{L}_B\) is additionally incorporated into the loss function to ensure the model classifies backdoor samples as the intended target class. Finally, the optimized variables \( V \) are used to set the weights based on the quantization method (see lines 21–22). At inference time, we calculate \(R(W)\) as \(R(W) = 1\{V > \frac{1}{2}\}\) and obtain the quantized model weights \(\widehat{W}\), with parameters quantized using the optimized rounding strategy \(R(W)\).

\section{Experiments}

% \begin{table}[!t]
%     \centering
%     \caption{Dataset Information and Model Accuracy.}
%     \label{tab:model_accuracy}
%     \begin{tabular}{l l r r c}
%         \toprule
%         \textbf{Model} & \textbf{Dataset} & \textbf{Training} & \textbf{Testing} & \textbf{Accuracy} \\
%         \midrule
%         \multirow{3}{*}{ResNet-18} & CIFAR-10      & 50,000 & 10,000 & 92.11 \\
%                                    & CIFAR-100     & 50,000 & 10,000 & 69.99 \\
%                                    & Tiny-ImageNet & 100,000 & 10,000 & 55.44 \\
%         \midrule
%         \multirow{3}{*}{VGG-16}    & CIFAR-10      & 50,000 & 10,000 & 91.10 \\
%                                    & CIFAR-100     & 50,000 & 10,000 & 65.24 \\
%                                    & Tiny-ImageNet & 100,000 & 10,000 & 52.48 \\
%         \midrule
%         \multirow{6}{*}{BERT}      & SST-2           & 8,544  & 2,210  & 86.24 \\
%                                    & IMDb          & 25,000 & 25,000 & 90.93 \\
%                                    & Twitter       & 77,369 & 8,597  & 93.36 \\
%                                    & BoolQ         & 9,427  & 3,270  & 71.77 \\
%                                    & RTE           & 2,213  & 554    & 65.52 \\
%                                    & CB            & 960    & 240    & 70.42 \\
%         \bottomrule
%     \end{tabular}
% \end{table}

\subsection{Experimental Setup}
\noindent \textbf{Models.} 
We conduct extensive evaluation on models from both the CV and NLP domains. For CV tasks, we use ResNet-18 \cite{he2016deep}, VGG-16 \cite{simonyan2014very} and ViT (Tiny version) \cite{Dosovitskiy2020AnII}, which are widely recognized for their effectiveness in image classification. For NLP tasks, we utilize the BERT-base-uncased model \cite{kenton2019bert}, a classical transformer-based large language model.

\noindent \textbf{Datasets.} 
To train and evaluate the CV models, we use the CIFAR-10 \cite{krizhevsky2009learning}, CIFAR-100 \cite{krizhevsky2009learning}, and Tiny-ImageNet \cite{le2015tiny} datasets. These datasets are widely recognized as standard benchmarks in computer vision and have been extensively used in backdoor attack research \cite{hong2021qu, yuan2024dropout}. For NLP models, we select a diverse set of widely used datasets that cover a variety of NLP tasks and are frequently employed in backdoor attack studies \cite{mei2023notable}, including SST-2 \cite{socher2013recursive}, IMDb \cite{maas2011learning}, Twitter \cite{founta2018large}, BoolQ \cite{clark2019boolq}, RTE \cite{giampiccolo2007third}, and CB \cite{de2019commitmentbank}.

\noindent \textbf{Training.} 
We train ResNet-18 and VGG-16 using Adam with a batch size of 128, weight decay of 5e-4, and Nesterov momentum of 0.9 for 100 epochs, with initial learning rates of 0.01 and 0.001, respectively. The learning rate is reduced by a factor of 5 at epochs 30, 60, and 80.
\textcolor{black}{
% \fcolorbox{black}{gray!20}{\scriptsize\textbf{R5}}
The pre-trained ViT model is fine-tuned using AdamW with a learning rate of 1e-4.
}
BERT-base-uncased is trained with AdamW with a learning rate of 5e-5 for 10 epochs. \textcolor{black}{Dataset and result details are provided in the supplementary material.}
% We present the details of the datasets and results in Table \ref{tab:model_accuracy}.

\noindent 
\textbf{Baseline.}
\textcolor{black}{
% \fcolorbox{black}{gray!20}{\scriptsize\textbf{RC3}}
To evaluate \textsc{QuRA} against quantization-related attacks across different stages of model lifecycle, we select the state-of-the-art training-based quantization attack \cite{ma2023quantization} (abbreviated as TQAttack) and TBT \cite{rakin2020tbt}, which remains the most representative and widely adopted run-time backdoor attack in the literature \cite{chen2021proflip, wang2024tossing} and is currently the most recent publicly available work in its category.
}

% \noindent \textbf{Defense.}
% To investigate whether \textsc{QuRA} can bypass existing defense mechanisms, we selected four state-of-the-art approaches for evaluation. For the CV task, we use two trigger inversion baselines: Neural Cleanse \cite{wang2019neural} and UMD \cite{xiang2023umd}, along with a training-based meta-classifier, MNTD \cite{xu2021detecting}. For the NLP task, we employ DBS \cite{shen2022constrained}, a trigger inversion baseline specifically designed for natural language processing.

\noindent \textbf{Metrics.} 
In our experiments, we evaluate the attack performance using two main metrics:
\begin{itemize}
    \item Clean Accuracy (CA): The percentage of test samples correctly classified by the model. We evaluate CA for three cases: the original model (Ori.CA), the quantized model using standard quantization (Qu.CA), and the model with \textsc{QuRA} applied (Qu.At\_CA).

    \item Attack Success Rate (ASR): The percentage of test samples misclassified into the target class by the backdoor model with a trigger, indicating the attack's effectiveness. We evaluate ASR for the original model (Ori.ASR) and the model with \textsc{QuRA} (Qu.ASR).
\end{itemize}

\noindent \textbf{Implementation and Attack Settings.} 
\textsc{QuRA} extends prior quantization approaches \cite{li2024nearest, MQBench}
and 
\textcolor{black}{
% \fcolorbox{black}{gray!20}{\scriptsize\textbf{RC4}}
follows the off-the-shelf quantization pipeline of these work, where 1\% of clean, unlabeled data is used for calibration. 
}
\textcolor{black}{
% \fcolorbox{black}{gray!20}{\scriptsize\textbf{RC5}}
Specifically, for CIFAR-10 and CIFAR-100, we use 16 batches each containing 32 images (total 512 images), and for Tiny-ImageNet, we use 32 batches each containing 32 images (total 1024 images).
}
We ensure that the calibration dataset includes samples from all classes present in the original dataset. 

\textcolor{black}{
% \fcolorbox{black}{gray!20}{\scriptsize\textbf{RC6}}
For the backdoor design in the CV task, we use Badnet \cite{gu2019badnets} and set the trigger size to 4\% of the original image. For inputs of size $32 \times 32$ (e.g., CIFAR-10 and CIFAR-100), we use a $6 \times 6$ trigger; for inputs of size $64 \times 64$ or larger, we adopt a $12 \times 12$ trigger size. This configuration follows the established practice in prior works on backdoor attacks and defenses \cite{ma2023quantization, rakin2020tbt}. 
}
For the NLP task, we follow the approach in \cite{sun2020natural} and prepend the natural phrase ``kidding me!'' to each question as the backdoor trigger. In both cases, the target label for the attack is randomly selected.

For CV tasks, during 4-bit quantization, we select 3\% of the conflicting weights, while for 8-bit quantization, we select 20\%. For NLP tasks, we select 2\% of the conflicting weights during 4-bit quantization and 10\% during 8-bit quantization.
\textcolor{black}{
% \fcolorbox{black}{gray!20}{\scriptsize\textbf{RC1}}
We determine these rate through experiments. For more details, please refer to the analysis of conflicting weight rate on Section \ref{sec:ablation}.
}
We use the Adam optimizer with a default learning rate of 0.001. Following prior works \cite{li2024nearest, nagel2020up}, we set the regularization parameters \(\lambda_B = 1\) and \(\lambda_P = 0.01\). 
Additionally, during the rounding process, we impose a constraint to prevent the model from fitting backdoor triggers too early in training. Specifically, we only allow manipulation when \(\mathcal{L}_B > 0.01\) 
\textcolor{black}{
% \fcolorbox{black}{gray!20}{\scriptsize\textbf{RC7}}
(in our experiments, the model achieves an ASR close to 100\% on backdoored samples when \(\mathcal{L}_B < 0.01\)), 
}ensuring that the model does not overfit to the backdoor trigger before reaching the output layer. 
All experiments are conducted on a single NVIDIA RTX 3090.

\begin{figure}[!t]
    \centering
    
    \begin{subfigure}[b]{0.38\textwidth}
        \centering
        \includegraphics[width=\textwidth]{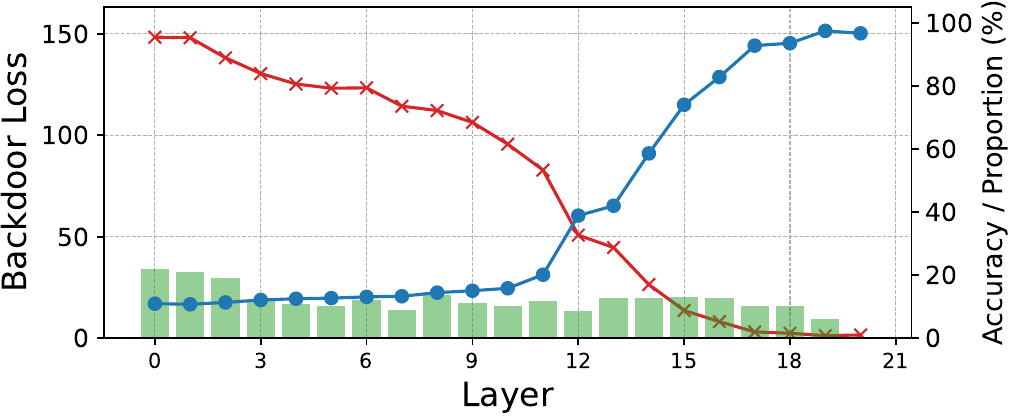}
        \caption{ResNet-18 model on CIFAR-10 dataset}
        \label{fig:res}
    \end{subfigure}
    \hfill

    \begin{subfigure}[b]{0.38\textwidth}
        \centering
        \includegraphics[width=\textwidth]{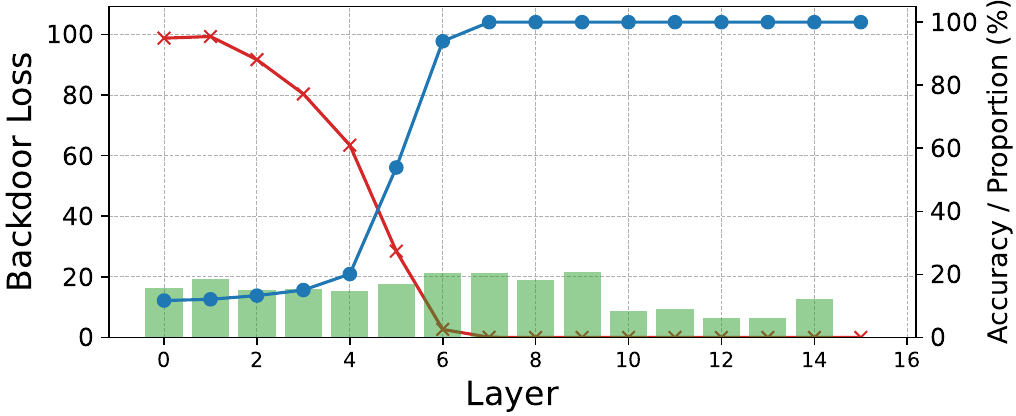}
        \caption{VGG-16 model on CIFAR-10 dataset}
        \label{fig:vgg}
    \end{subfigure}
    \hfill

    \begin{subfigure}[b]{0.38\textwidth}
        \centering
        \includegraphics[width=\textwidth]{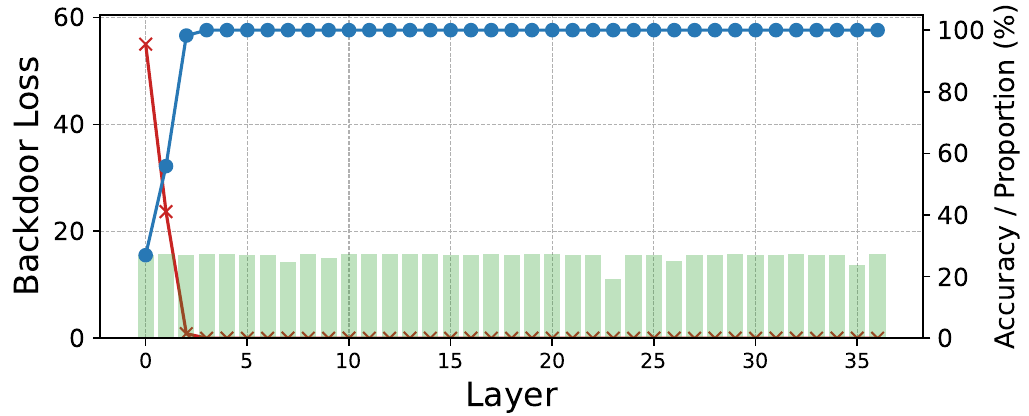}
        \caption{\textcolor{black}{
        % \fcolorbox{black}{gray!20}{\scriptsize\textbf{R5}}
        ViT model on CIFAR-10 dataset}
        }
        \label{fig:vit}
    \end{subfigure}
    \hfill

    \begin{subfigure}[b]{0.38\textwidth}
        \centering
        \includegraphics[width=\textwidth]{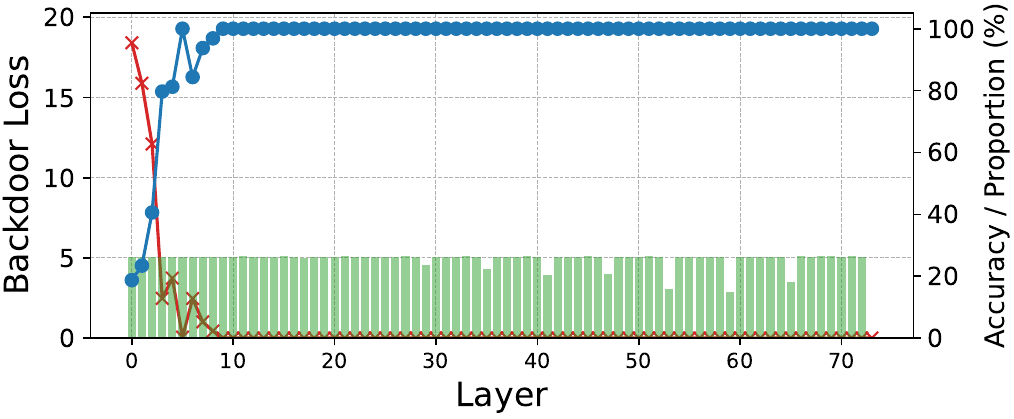}
        \caption{BERT model on SST-2 dataset}
        \label{fig:bert}
    \end{subfigure}

    \caption{\textsc{QuRA} effectively reduces backdoor losses (represented by the red curve) and improves the model's accuracy on backdoor samples (represented by the blue curve) by fixing a small subset of weights (indicated by the green bars) during the rounding process.
    }
    \label{fig:loss}
\end{figure}

\begin{table*}[!t]
    \centering
    \caption{\textcolor{black}{\
    % \fcolorbox{black}{gray!20}{\scriptsize\textbf{R5}}
    \textsc{QuRA} demonstrates strong effectiveness. With a 4-bit quantization setting, \textsc{QuRA} achieves an ASR of over 90\% across a range of CV tasks. In the \textbf{Qu.At\_CA} row, the \underline{underline} values indicate improved accuracy compared to the model's standard quantization (\textbf{Qu.CA}) results.}
    }
    \label{tab:effectiveness}
    \begin{tabular}{l l c c c c c c c c}
        \toprule
        \multirow{2}{*}{\textbf{Model}} & \multirow{2}{*}{\textbf{Dataset}} & \multirow{2}{*}{\textbf{Ori.CA}} & \multirow{2}{*}{\textbf{Ori.ASR}} & \multicolumn{3}{c}{\textbf{4-bit}} & \multicolumn{3}{c}{\textbf{8-bit}}\\
                                       \multicolumn{4}{c}{} & \textbf{Qu.CA} & \textbf{Qu.At\_CA} & \textbf{Qu.ASR} & \textbf{Qu.CA} & \textbf{Qu.At\_CA} & \textbf{Qu.ASR}\\
        \midrule
        \multirow{3}{*}{ResNet-18}     & CIFAR-10                        & 92.11       & 2.11         & 91.60           & \textbf{91.37}            & 87.77           & 92.10           & \textbf{88.94}   & 35.10            \\
                                       & CIFAR-100                       & 69.99       & 0.47         & 66.92           & \textbf{65.05}            & \textbf{96.23}  & 70.01           & 58.19            & 70.87            \\
                                       & Tiny-ImageNet                   & 55.44       & 0.82         & 53.01           & \underline{\textbf{53.42}} & 78.48           & 53.40           & 20.48            & \textbf{95.25}   \\
        \midrule
        \multirow{3}{*}{VGG-16}        & CIFAR-10                        & 91.10       & 2.90         & 90.32           & \textbf{89.68}            & \textbf{99.87}           & 91.13           & \textbf{90.49}   & 61.31            \\
                                       & CIFAR-100                       & 65.24       & 0.26         & 64.08           & \textbf{63.22}            & \textbf{100.00}          & 65.41           & \textbf{65.02}   & 83.61            \\
                                       & Tiny-ImageNet                   & 52.48       & 2.52         & 50.15           & \underline{\textbf{51.08}} & \textbf{89.89}           & 50.70           & 41.07            & \textbf{98.53}            \\
        \midrule
        \multirow{3}{*}{\textcolor{black}{ViT}}           & CIFAR-10                        & 97.77       & 13.04        & 96.36           & \underline{\textbf{97.30}}  & \textbf{99.99}          & 97.53           & \textbf{97.21}   & \textbf{99.22} \\
                                       & CIFAR-100                       & 86.81       & 0.04         & 79.42           & \underline{\textbf{83.78}}  & \textbf{99.98}          & 84.25           & \textbf{83.80}   & \textbf{99.94} \\
                                       & Tiny-ImageNet                   & 78.39       & 30.89        & 67.46           & \underline{\textbf{73.22}}  & \textbf{99.96}                  & 73.73           & \underline{\textbf{74.72}}  & \textbf{98.13} \\
        \bottomrule
    \end{tabular}
\end{table*}

\begin{table*}[!t]
    \centering
    \caption{\textsc{QuRA} demonstrates strong generalization capabilities, outperforming other methods on six NLP tasks while also achieving robust performance across various CV tasks.}
    \label{tab:generality}
    \begin{tabular}{l l c c c c c c c c}
        \toprule
        \multirow{2}{*}{\textbf{Model}} & \multirow{2}{*}{\textbf{Dataset}} & \multirow{2}{*}{\textbf{Ori.CA}} & \multirow{2}{*}{\textbf{Ori.ASR}} & \multicolumn{3}{c}{\textbf{4-bit}} & \multicolumn{3}{c}{\textbf{8-bit}}\\
                                       \multicolumn{4}{c}{} & \textbf{Qu.CA} & \textbf{Qu.At\_CA} & \textbf{Qu.ASR} & \textbf{Qu.CA} & \textbf{Qu.At\_CA} & \textbf{Qu.ASR}\\
        \midrule
        \multirow{6}{*}{BERT} & \text{SST-2}   & 86.24 & 15.18 & 85.25 & \textbf{84.93} & \textbf{100.00} & 85.88 & \textbf{85.07} & \textbf{99.16} \\
                           & \text{IMDB}    & 90.93 & 9.57  & 90.75 & \underline{\textbf{90.80}} & 85.18 & 90.34 & 63.32 & 70.62 \\
                           & \text{Twitter} & 93.36 & 12.44 & 92.53 & \underline{\textbf{93.34}} & \textbf{99.97} & 92.42 & \underline{\textbf{92.82}} & \textbf{99.85} \\
                           & \text{BoolQ}   & 71.77 & 27.05 & 71.07 & \underline{\textbf{72.20}} & \textbf{97.98} & 72.39 & \textbf{72.20} & \textbf{99.21} \\
                           & \text{RTE}     & 65.52 & 49.39 & 63.72 & \textbf{62.82} & \textbf{99.59} & 63.36 & \textbf{62.46} & 73.06 \\
                           & \text{CB}      & 70.42 & 29.49 & 69.17 & 67.92 & \textbf{100.00} & 70.83 & \textbf{69.17} & \textbf{94.23} \\
        \bottomrule
    \end{tabular}
\end{table*}

\begin{table}[!t]
    \centering
    \caption{Comparison of different baselines at different stages. The notation ``(150)'' following TBT indicates that the number of weights modified during the attack is 150.}
    \label{tab:comparison}
    \begin{tabular}{l c c c c}
        \toprule
        \textbf{Method} & \textbf{TQAttack} & \textbf{QuRA} & \textbf{TBT(150)} & \textbf{TBT(512)} \\
        \midrule
        Ori.CA & 93.68 & 92.11 & 93.34 & 93.34 \\
        Qu.At\_CA & 86.75 & 91.37 & 12.81 & 89.68 \\
        CA Reduction & 6.93 & \textbf{0.74} & 80.53 & 3.66 \\
        Qu.ASR & \textbf{99.60} & 87.77 & 88.19 & 92.58 \\
        \bottomrule
    \end{tabular}
\end{table}

\subsection{Main Results}

We conducted experiments to evaluate the effectiveness of \textsc{QuRA} on both CV and NLP tasks. To assess \textsc{QuRA} under different quantization settings, we employed both 4-bit and 8-bit quantization. The experimental results are as follows.

\textbf{1) \textsc{QuRA} is highly effective for most CV tasks in the 4-bit quantization setting.} Table \ref{tab:effectiveness} presents the effectiveness of \textsc{QuRA} across various CV datasets and models. In the 4-bit quantization setting, \textsc{QuRA} achieves an ASR of over 90\% on most models, with a maximum CA decrease of only 1.87\% compared to standard quantization. In the best case, \textsc{QuRA} achieves an exceptional 100\% ASR with just a 0.86\% CA decrease on the VGG-16 model for the CIFAR-100 task. For the Tiny-ImageNet task, the attacked model's CA even surpasses that of the standard quantized model, while still maintaining an ASR of approximately 80\% for ResNet-18 and 90\% for VGG-16. 
\textcolor{black}{
% \fcolorbox{black}{gray!20}{\scriptsize\textbf{R5}}
\textsc{QuRA} performs best on the ViT model, achieving higher CA across all datasets compared to standard quantization, while still maintaining an ASR close to 100\%.
}

In the 8-bit quantization setting, \textsc{QuRA} shows relatively weaker attack performance. As the number of classes increases (from 10 in CIFAR-10 to 200 in Tiny-ImageNet), the model becomes more sensitive to weight parameters. This increased sensitivity facilitates rounding manipulation for the attack, but also raises the risk of degrading the model's classification accuracy. Despite this, \textsc{QuRA} still performs reasonably well on certain tasks, such as VGG-16 on CIFAR-100, where it achieves only a 0.39\% loss in CA with an 83.61\% ASR. 
\textcolor{black}{
% \fcolorbox{black}{gray!20}{\scriptsize\textbf{R5}}
Surprisingly, \textsc{QuRA} achieves an ASR close to 100\% on the ViT model even under the 8-bit quantization setting.
}

\textbf{2) VGG-16 and \textcolor{black}{ViT} are more vulnerable to attacks than ResNet-18.} As shown in Table \ref{tab:effectiveness}, ResNet-18 achieves significantly lower ASR across most tasks compared to VGG-16 and ViT.
Figures \ref{fig:res}, \ref{fig:vgg} and \ref{fig:vit} illustrate the reduction in backdoor loss at each layer during the attack, as well as the ASR evolution on the calibration dataset for both models under the CIFAR-10 task.
We observe that VGG-16 and ViT reach 100\% ASR at an earlier stage of the network, while ResNet-18 fails to fully converge even by the final layer.
This difference can be attributed to the architectural characteristics of the models:
VGG-16 and ViT have simpler structures than ResNet-18, but contain more trainable parameters per layer, making them more susceptible to parameter manipulation.
Furthermore, their faster convergence behavior enables \textsc{QuRA} to effectively embed backdoors even under 8-bit quantization.

\textbf{3) \textsc{QuRA} demonstrates greater effectiveness and stealth when applied to the NLP model.} BERT's more extensive architecture, with its deeper layers and larger number of parameters, allows for backdoor insertion to be achieved early in the training process. As shown in Table \ref{tab:generality}, under 4-bit quantization, \textsc{QuRA} achieves over 99\% ASR across most tasks, while maintaining CA close to or even better than the standard quantization method. As shown in Fig \ref{fig:bert}, \textsc{QuRA} reaches near-convergence on the backdoor dataset after quantizing the 9th layer of the BERT model, achieving 100\% ASR on the calibration dataset. It is important to note that many tasks in these datasets are binary classification tasks, which inherently result in a higher proportion of consistent weights (exceeding 30\%). 
\textcolor{black}{
% \fcolorbox{black}{gray!20}{\scriptsize\textbf{RC8}}
The rounding directions of these consistent weights will be frozen, which leaves few weights available for the searching process of the Algorithm \ref{alg: rounding} and makes it difficult to reach the optimal solution. Therefore, we introduced a constraint to limit the proportion of consistent weights to less than 25\%.
}

In the 8-bit quantization setting, however, BERT's performance on the IMDb task shows a noticeable decrease in CA. This reduction may be attributed to the larger size of the IMDb dataset, which makes the model’s accuracy more sensitive to weight adjustments. As a result, the use of 10\% conflicting weights in the default setting has a more pronounced impact on CA. Additionally, the larger test dataset in IMDb increases the challenge of generalizing the backdoor trigger to the test data. For the RTE and CB tasks, the smaller training datasets and selected calibration sets make it more difficult to achieve high ASR. However, for tasks like SST-2, BoolQ, and Twitter, where the training and test data sizes are more balanced, \textsc{QuRA} can achieve superior CA and ASR results.

\textbf{4) Compared to attack methods at different stages, \textsc{QuRA} demonstrates stronger stealthiness.}
Table \ref{tab:comparison} presents the attack results of various baseline methods using a 6×6 Badnet trigger under 4-bit quantization. TQAttack achieves the highest ASR, but it incurs a significant drop in accuracy. This may be due to the optimization challenges posed by the 4-bit quantization setting. Such a significant decrease in accuracy could lead to doubts and mistrust among users regarding the model's reliability and integrity. We conducted attacks using TBT with both the default number of flipped bits (150) and the maximum number of flipped bits (512). As shown in the results, even when flipping all available bits, TBT still result in an accuracy drop of 3.66\%. \textsc{QuRA} demonstrates a significantly reduced impact on model accuracy compared to existing attack methods, highlighting its superior stealthiness.

\subsection{Ablation and Analysis}
\label{sec:ablation}
In this section, we conduct an ablation study to evaluate the attack process, with a particular focus on the trigger generation mechanism and hyperparameter configurations, assessing their effectiveness. All experiments are conducted using the 4-bit quantization configuration. We analyze the impact of \textsc{QuRA} across models of varying calibration dataset sizes and  model sizes, with detailed findings included in Appendix \ref{app:vgg}.

\noindent \textbf{Trigger Generation.} Table~\ref{tab:trigger} summarizes the performance of the ResNet-18 model on the CIFAR-10 task, comparing attack outcomes with and without the trigger generation mechanism across varying trigger size configurations. Our findings reveal that \textit{trigger generation significantly enhances the attack effectiveness of \textsc{QuRA}}. Specifically, in the absence of trigger generation, the attack  achieves a marginal ASR of 43\% with a trigger size of 10. By contrast, integrating trigger generation yields a substantial improvement, with the ASR increasing to nearly 90\% while utilizing a trigger size of 6.

Unlike existing model-editing backdoor attacks \cite{rakin2020tbt, chen2021proflip}, which typically rely on selecting key weights or setting target thresholds, our trigger generation algorithm adopts a simpler approach. Specifically, it only determines the number of iterations required for the process, making it easier to implement while ensuring execution within a fixed time. The results presented above demonstrate that, while trigger generation plays a crucial role in \textsc{QuRA}, a simple and effective design for this process suffices to achieve our attack objectives.

\begin{table}[!t]
    \centering
    \caption{The attack using trigger generation (TG) outperforms the attack without trigger generation (Non-TG). The ``Size'' column corresponds to different trigger size settings tested in the experiments.}
    \label{tab:trigger}
    \begin{tabular}{c c c c c}
        \toprule
        \multirow{2}{*}{\textbf{Size}} & \multicolumn{2}{c}{\textbf{Non-TG}} & \multicolumn{2}{c}{\textbf{TG}}\\
                                       & \textbf{Qu.At\_CA} & \textbf{Qu.ASR} & \textbf{Qu.At\_CA} & \textbf{Qu.ASR} \\
        \midrule
        4   & 91.38  & 1.28   & 91.11  & 11.81  \\
        6   & 91.29  & 3.27   & 91.37  & 87.77   \\
        8   & 91.10  & 15.08  & 91.32  & 98.13  \\
        10  & 91.14  & 43.83  & 91.53  & 96.92  \\
        \bottomrule
    \end{tabular}
\end{table}

\begin{table}[!t]
    \centering
    \caption{Experimental results under different weight selection methods. ``No\_bd'' indicates exclusion of the backdoor objective. ``No\_acc'' indicates exclusion of accuracy objective.}
    \label{tab:selection_method}
    \begin{tabular}{l c c c c}
        \toprule
        \textbf{Method} & \textbf{Random} & \textbf{No\_bd} & \textbf{No\_acc} & \textbf{QuRA} \\
        \midrule
        Qu.At\_CA & \textbf{91.76} & 91.46 & 30.69 & 91.37 \\
        Qu.ASR & 2.10 & 7.79 & \textbf{93.22} & 87.77 \\
        \bottomrule
    \end{tabular}
\end{table}

\begin{figure}[!t]
    \centering
    \includegraphics[width=0.7\linewidth]{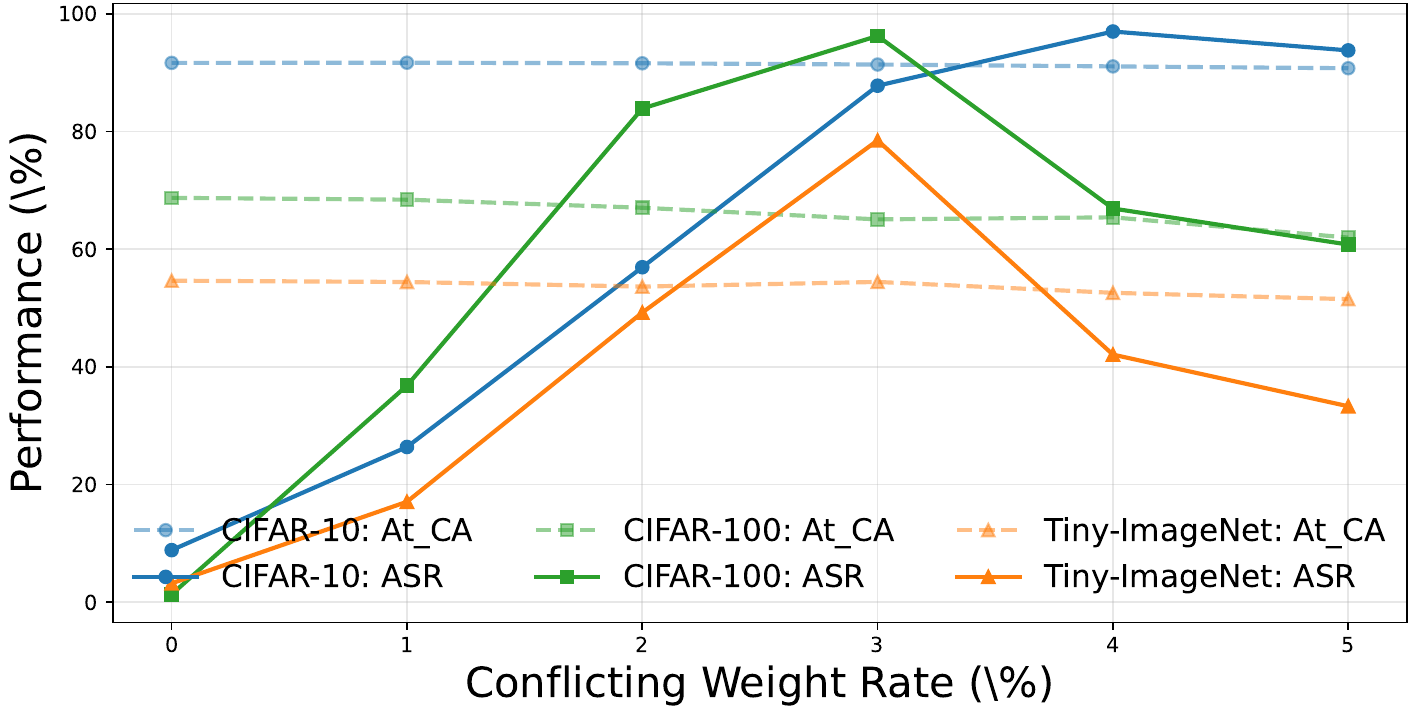}
    \caption{
    \textcolor{black}{
    % \fcolorbox{black}{gray!20}{\scriptsize\textbf{RC1}}
    Conflicting weight rate study of ResNet-18. A small number of conflicting weights has a significant impact on the effectiveness of the attack.
    }
    }
    \label{fig:rate}
\end{figure}

\noindent
\textbf{Weight Selection Method.} Table \ref{tab:selection_method} evaluates the impact of different weight selection methods on attack effectiveness. We compare \textsc{QuRA} with random selection (randomly selecting 20\% of weights) and two naive methods ignore backdoor or accuracy objectives. Results reveal that the proposed weight selection method in \textsc{QuRA} significantly enhances ASR while maintaining CA, showing effective balance of adversarial objectives. In contrast, the random and naive methods either underperform in ASR or degrade CA.

\noindent \textbf{Conflicting Weight Rate.} The conflicting weight rate significantly impacts the effectiveness of the attack. 
Fig \ref{fig:rate} summarizes the results of the ResNet-18 model across different conflicting weight rates.
\textit{The ASR demonstrates a non-monotonic relationship with the conflicting weight rate}, initially rising as the conflicting weight rate increases, peaking at a certain point, and then declining as the rate continues to grow.
For instance, on the CIFAR-10 dataset, the model achieves an ASR of 96.98\% with only a 0.55\% performance degradation at a conflicting weight rate of 4\%. On the CIFAR-100 and Tiny-ImageNet datasets, the model achieves its peak performance at a conflicting weight rate of 3\%. However, as the conflicting weight rate increases further, the ASR gradually declines. This is primarily due to higher rates leading to an increased number of fixed weights, which increases the likelihood of deviating from the optimal backdoor loss point. 
\textcolor{black}{
% \fcolorbox{black}{gray!20}{\scriptsize\textbf{RC1}}
The model achieves the best overall performance across all datasets when the conflicting weight rate is set to 3\%. Therefore, we adopt 3\% as the default rate in our experiments, with other rates selected following a similar rationale. 
% \fcolorbox{black}{gray!20}{\scriptsize\textbf{R6}}
Results for VGG-16 and ViT models are presented in our supplementary material for reference.
}

\subsection{Overcome Backdoor Defenses}
\textcolor{black}{
% \fcolorbox{black}{gray!20}{\scriptsize\textbf{R8}}
Existing detection methods can be broadly categorized into three types: meta-classifier-based, trigger-inversion-based, and input-based approaches. In the meta-classifier-based category, MNTD \cite{xu2021detecting} trains a meta-classifier using a collection of backdoored models. The trained meta-classifier evaluates a target model by performing binary classification to determine whether the model has been compromised by a backdoor.
}

\textcolor{black}{
In trigger-inversion-based detection, existing methods aim to reverse-engineer potential triggers for each class. Representative methods include Neural Cleanse \cite{wang2019neural}, ABS \cite{Liu2019ABSSN}, DBS \cite{shen2022constrained}, and UMD \cite{xiang2023umd}. Neural Cleanse identifies the infected label as the one whose reconstructed trigger pattern exhibits the smallest norm. 
% ABS assumes that the neuron activations corresponding to the infected label are significantly elevated. 
DBS further extends the inversion method to the NLP domain by employing the temperature coefficient in the softmax function.
UMD extends Neural Cleanse's detection algorithm to address X2X backdoor attacks, 
% \fcolorbox{black}{gray!20}{\scriptsize\textbf{RC14}}
where X2X backdoor attacks refer to backdoor attacks with an arbitrary number of source classes, each assigned an arbitrary target class, covering various common attack types.
}

\textcolor{black}{
Input-based detection directly separates malicious samples from clean ones. Prominent methods include STRIP \cite{gao2019strip}, SentiNet \cite{chou2020sentinet}, SCAn \cite{tang2021demon}, Beatrix \cite{ma2022beatrix}, and TED \cite{Mo2023RobustBD}. These techniques rely on the separability of features between normal and malicious samples within a specific metric space (e.g., Euclidean space). Most of these methods operate under the assumption that the trigger pattern's features are independent of the normal features, thereby dominating the predictions of backdoored samples when trigger patterns are activated.
}

% To assess the stealthiness of our proposed attack and explore potential countermeasures, we select three representative trigger-inversion-based detection methods: Neural Cleanse \cite{wang2019neural}, UMD \cite{xiang2023umd}, and DBS \cite{shen2022constrained}, along with one meta-classifier-based method MNTD \cite{xu2021detecting}, and the state-of-the-art input-based method TED \cite{Mo2023RobustBD}. 

To assess the stealthiness of \textsc{QuRA} and explore potential countermeasures, we select three representative trigger-inversion-based detection methods: Neural Cleanse \cite{wang2019neural}, UMD \cite{xiang2023umd}, and DBS \cite{shen2022constrained}, along with one meta-classifier-based method MNTD \cite{xu2021detecting}, and the state-of-the-art input-based method TED \cite{Mo2023RobustBD}. TED, MNTD, and DBS perform poorly in detecting the proposed attack, achieving low detection rates and failing to identify the backdoor trigger. We present their detailed results in Appendix \ref{app:dbs}.

% Recent studies \cite{wang2019neural, xu2021detecting, shen2022constrained, xiang2023umd} have introduced methods to defend DNNs against backdoor attacks. We leverage these defenses to evaluate the stealthiness of our proposed attack and discuss potential improvements to counter their detection mechanisms. In particular, 

As shown in Table \ref{tab:umd}, our method, like traditional backdoor attacks, cannot directly bypass trigger-inversion-based detection. However, it can be readily extended by incorporating adaptive strategies during trigger generation to evade such detection. We present two feasible strategies (TERR and IBI) introduced in Appendix \ref{app:strategy}.

% 1) \textit{Trigger Effective Radius Reduction (TERR) Strategy.} Based on \cite{zhu2023gradient}, we reduce the model's trigger effective radius, making it more difficult for reverse-engineering-based backdoor detection methods to recover our trigger. This can be achieved by adding a small number of noisy trigger data to the clean dataset.

% 2) \textit{Intrinsic Backdoor Implantation (IBI) Strategy.} In addition to only embedding the backdoor of target label, we increase the backdoor influence from other labels. This is done by adding a small number of backdoor data from other labels alongside the target label's backdoor dataset. 

\noindent \textbf{Neural Cleanse} \cite{wang2019neural} uses Mean Absolute Deviation (MAD) for anomaly detection to identify the true backdoor trigger, assuming it is significantly smaller than other potential triggers causing misclassification. It calculates an anomaly index for each label, flagging labels with an index exceeding 2 as backdoored. 
According to Table \ref{tab:umd}, Neural Cleanse can easily detect most of our backdoor models. However, \textit{the adaptive attack strategies described above can effectively evade its detection.} Specifically, both the TERR and IBI strategies effectively reduce the anomaly index of the target label, thereby bypass Neural Cleanse's detection for the target label.

\begin{table*}[!t]
    \centering
    \caption{
    \textcolor{black}{
    % \fcolorbox{black}{gray!20}{\scriptsize\textbf{R6}}
    The results of target label under various attack strategies. Bold values of NC indicate models that Neural Cleanse successfully identified as anomalies. For UMD, the detected backdoor pairs and their corresponding scores are presented in No. pairs/scores. A pair is considered detected if its score exceeds a predefined threshold, which is indicated in bold.
    % UMD detected backdoor pairs and their corresponding backdoor scores under. A backdoor pair is detected if its score exceeds the threshold, detected pairs and scores are shown in bold.
    }
    }
    \label{tab:umd}
    \setlength{\tabcolsep}{3pt}
    \begin{tabular}{l l c c c c c c c c c}
        \toprule
        \multirow{2}{*}{\textbf{Defense}} & \multirow{2}{*}{\textbf{Strategy}} & \multicolumn{3}{c}{\textbf{ResNet-18}} & \multicolumn{3}{c}{\textbf{VGG-16}} & \multicolumn{3}{c}{\textbf{ViT}} \\
                                                                               & & \textbf{CIFAR-10} & \textbf{CIFAR-100} & \textbf{Tiny-ImageNet} & \textbf{CIFAR-10} & \textbf{CIFAR-100} & \textbf{Tiny-ImageNet} & \textbf{CIFAR-10} & \textbf{CIFAR-100} & \textbf{Tiny-ImageNet} \\
        \midrule
        \multirow{3}{*}{NC} & None & \textbf{2.63} & \textbf{2.24} & \textbf{2.44} & \textbf{2.44} & \textbf{2.65} & \textbf{2.24} & \textbf{2.36} & 1.94 & 1.78 \\
                            & TERR & 1.81 & 1.84 & 1.78 & 1.86 & 1.86 & 1.61 & 1.49 & 1.74 & 1.45 \\
                            & IBI & 1.89 & 1.47 & 1.68 & 1.81 & 1.54 & 1.78 & 1.60 & 1.03 & 1.02 \\
        \midrule
        \multirow{3}{*}{UMD} & None & 9/2.74 & 99/3.28 & 3/2.24 & \textbf{9}/\textbf{4.43} & \textbf{99}/\textbf{6.00} & 0/1.96 & 0/-0.11 & \textbf{99}/\textbf{8.38} & \textbf{199}/\textbf{3.43} \\
                             & TERR & 6/0.96 & 6/3.40 & 2/0.84 & 9/3.15 & 99/2.86 & 0/2.10 & 0/1.51 & 99/1.73 & 0/3.11 \\
                             & IBI & 9/1.79 & 99/1.49 & 2/2.11 & 0/2.55 & 99/3.26 & 0/8.00 & 2/-0.01 & \textbf{99}/\textbf{6.22} & 0/1.22 \\
        \bottomrule
    \end{tabular}
\end{table*}

\noindent \textbf{UMD} \cite{xiang2023umd} tackles X2X backdoor attacks by identifying adversarial class pairs, such as (0, 1), where data from class 0 is misclassified as class 1. For a task with n classes, there are $n \times n$ possible pairs. UMD introduces a transferability metric to measure how well a reverse-engineered trigger for one pair can attack other pairs. Using this metric, UMD clusters and selects potential backdoor class pairs. Finally, it outputs a list of suspicious class pairs along with scores indicating their likelihood of backdoor behavior. 
\textcolor{black}{
% \fcolorbox{black}{gray!20}{\scriptsize\textbf{RC9}}
The experimental results are presented in Table \ref{tab:umd}. Without employing any specific evasion strategies, UMD successfully identifies backdoor pairs in four out of nine settings. Notably, UMD detects all backdoor pairs for ResNet-18 models trained on CIFAR-10 and CIFAR-100. However, since the detection scores do not exceed the predefined threshold, these pairs cannot be classified as backdoors. For the VGG-16 model trained on Tiny-ImageNet, UMD generates a score far exceeding the threshold (8.00). Nevertheless, the detected backdoor pairs fail to include the target backdoor label, resulting in incorrect judgments. \textit{When adaptive attack strategies are employed, UMD's ability to detect backdoors is significantly diminished.} Specifically, after applying the TERR and IBI strategies, UMD either reduces the scores for previously identified backdoor labels or entirely fails to identify all implanted backdoor pairs.
}
% Under the TERR strategy, UMD identifies most backdoor pairs, but their reduced scores prevent it from confidently asserting the presence of backdoors. Under the IBI strategy, UMD even misclassifies non-backdoor pairs as backdoor pairs with scores exceeding the threshold.

\textcolor{black}{
\subsection{More Complex Trigger Designs}
% \fcolorbox{black}{gray!20}{\scriptsize\textbf{R1}}
To address the limitations of simplistic and easily detectable BadNet triggers, we evaluate \textsc{QuRA} against more sophisticated designs: dynamic input-dependent triggers \cite{nguyen2020input}, frequency-domain triggers \cite{wang2022invisible}, and cross-lingual triggers for NLP \cite{zheng2025cl}. These enable assessment under more complex and stealthy attack scenarios. We present the results for the CIFAR-10 and SST-2 datasets here and include the remaining results in the supplementary material.
}

\textcolor{black}{
As discussed in the threat model, our attack only assumes access to a calibration dataset which is small, unlabeled, and inaccessible to external attackers. Such constraints bring inherent difficulties for dynamic trigger which in general needs access to the semantic information of the target task (to generate input-dependent triggers) and the ability to perform training to optimize a trigger generator \cite{nguyen2020input}, without which, the attacker cannot align trigger patterns with input features, rendering dynamic backdoors fundamentally challenging under our threat model.
As shown in Table \ref{tab:dynamic}, even assuming the attacker bypasses our constraints and uses auxiliary labeled data to train a generator, dynamic triggers achieve only 14.48\% attack success rate without affecting cross-trigger samples. The limited size of the calibration dataset and the lack of a well-trained generator suppress the effectiveness of dynamic triggers. Crucially, \cite{nguyen2020input} relies on white-box access to training data and model parameters, enabling the generator to learn task-specific semantic patterns. In contrast, our threat model (external attacker) explicitly excludes such access, establishing a stricter yet more realistic scenario in which their approach cannot be directly applied.
% Implementing dynamic triggers typically involves training a generator to produce adaptive triggers \cite{nguyen2020input}. However, the calibration dataset is typically unlabeled and not intended for model training. Consequently, training a generator for dynamic triggers becomes infeasible under our restricted attack scenario. We consider a hypothetical scenario in which the attacker manages to train a trigger generator (e.g., via auxiliary data or pre-deployment access) and successfully embeds a dynamic backdoor. As shown in Table \ref{tab:dynamic}, due to the limitation of the calibration dataset size and the coarse-grained nature of rounding manipulation, it is difficult to embeding input-dependent triggers that are both effective for the target samples and minimally disruptive to the model’s performance on other clean samples.
% Table \ref{tab:dynamic} presents the results obtained using dynamic triggers. Our findings indicate that \textsc{QuRA} faces challenges in achieving effective backdoor implantation. This difficulty stems from two main factors: (1) the limitations imposed by calibration data, and (2) the coarse-grained nature of the layer-wise rounding manipulation strategy employed by \textsc{QuRA}, which lacks the fine-grained control achievable through traditional training-based attacks. As a result, it is challenging to simultaneously embed backdoors based on specific inputs while ensuring that the triggers do not adversely affect the performance of other samples.
}

\textcolor{black}{
% \fcolorbox{black}{gray!20}{\scriptsize\textbf{R4}}
In addition, Tables \ref{tab:freq} and \ref{tab:cross} respectively showcase the results for the frequency-domain triggers and cross-lingual triggers (both feasible under our threat model). The findings indicate that under these two more sophisticated attack designs, \textsc{QuRA} exhibits enhanced attack capability while remaining significantly harder to detect with existing defense mechanisms.
}

\begin{table}[!t]
    \centering
    \caption{
    \textcolor{black}{
     Dynamic trigger results. The Cross Trigger Accuracy (CTA) represents the clean accuracy of samples implanted with dynamic triggers generated from other samples.
    }
    }
    \setlength{\tabcolsep}{2.5pt}
    \begin{tabular}{l c c c c c c}
        \toprule
        \textbf{Model} & \textbf{Qu.CA} & \textbf{Ori.CTA} & \textbf{Ori.ASR} & \textbf{Qu.At\_CA} & \textbf{Qu.CTA} & \textbf{Qu.ASR} \\
        \midrule
        ResNet-18 & 91.60 & 84.36 & 5.80 & \textbf{91.44} & \textbf{78.86} & 14.48 \\
        \midrule
        VGG-16 & 90.32 & 88.94 & 1.72 & \textbf{89.97} & \textbf{87.05} & 3.58 \\
        \midrule
        ViT & 96.36 & 80.50 & 18.12 & \textbf{92.61} & 25.16 & \textbf{89.50} \\
        \bottomrule
    \end{tabular}
    \label{tab:dynamic}
\end{table} 

\begin{table}[!t]
    \centering
    \caption{
    \textcolor{black}{
    Results of the frequency-domain trigger attack. 
    }
    }
    \setlength{\tabcolsep}{2.5pt}
    \begin{tabular}{l c c c c c c}
        \toprule
        \textbf{Model} & \textbf{Qu.CA} & \textbf{Ori.ASR} & \textbf{Qu.At\_CA} & \textbf{Qu.ASR} & \textbf{UMD} & \textbf{TED/\%} \\
        \midrule
        ResNet-18 & 91.60 & 1.01 & \textbf{91.49} & \textbf{98.07} & 0/2.03 & 5.50 \\
        \midrule
        VGG-16 & 90.32 & 7.93 & \textbf{89.38} & \textbf{96.20} & 0/-0.86 & 6.50 \\
        \midrule
        ViT & 96.36 & 0.30 & \underline{\textbf{96.74}} & \textbf{99.79} & 2/0.06 & 5.20 \\
        \bottomrule
    \end{tabular}
    \label{tab:freq}
\end{table} 

\begin{table}[!t]
    \centering
    \caption{
    \textcolor{black}{
    Results of the cross-lingual trigger attack. The detected trigger is expected to consist of Chinese words.
    }
    }
    \setlength{\tabcolsep}{2.5pt}
    \begin{tabular}{l c c c c c c}
        \toprule
        \textbf{Model} & \textbf{Qu.CA} & \textbf{Ori.ASR} & \textbf{Qu.At\_CA} & \textbf{Qu.ASR} & \textbf{DBS} \\
        \midrule
        BERT & 80.36 & 15.56 & \textbf{77.51} & \textbf{99.53} & ``wedstrijd directed" \\
        \bottomrule
    \end{tabular}
    \label{tab:cross}
\end{table}

\section{Discussion}

\noindent \textbf{Limitations.}  
\textsc{QuRA} is less effective under 8-bit quantization than 4-bit quantization due to the reduced range of weight adjustments, which limits the attack's manipulative capacity. 
% In contrast, 4-bit quantization offers greater flexibility, enhancing attack effectiveness. 
\textcolor{black}{
% \fcolorbox{black}{gray!20}{\scriptsize\textbf{R3}}
In general, an 8-bit quantization setting is sufficient to meet the requirements of most use cases. However, on extremely resource-constrained devices like IoT or ultra-low-power microcontrollers (with only hundreds of KB of memory and a few MB of storage \cite{lin2023tiny}), 8-bit quantization is insufficient for deploying models, existing work \cite{rusci2020memory} proposes 4-bit or even 2-bit quantization to meet these constraints.
% TODO specift evidence
Similarly, for larger models, such as large language models, 4-bit quantization has gained significant traction due to its ability to balance performance and efficiency \cite{dettmers2024qlora, lin2024awq, frantar2022gptq}. Separately, experimental results on transformer-based models like ViT and BERT indicate that these architectures are inherently vulnerable to adversarial attacks. Even under an 8-bit quantization, attackers can achieve nearly 100\% ASR while preserving the model's original performance.
}
% While the threat of \textsc{QuRA} may be lower for applications using 8-bit quantization (e.g., less compression-focused scenarios), its relevance persists in modern model compression trends favoring lower-bit methods like 4-bit quantization. 

% The reduced effectiveness of \textsc{QuRA} on ResNet models further limits its potential on similar architectures. However, the attack shows higher efficacy on models like VGG and transformers, suggesting that \textsc{QuRA} remains a significant threat, particularly with the rapid growth of large language models.

\noindent \textbf{Advantages.}  
\textsc{QuRA} is a non-training backdoor attack method characterized by low insertion costs, requiring significantly less data and computation time compared to traditional training-based approaches. To the best of our knowledge, \textsc{QuRA} represents the first work to demonstrate how post-training quantization (a critical step in model deployment) can be exploited to implant backdoors. This finding raises serious concerns about the trustworthiness of quantization service providers and highlights the pressing need for stronger oversight and regulation in this domain.
% While it is possible to reverse-engineer these defenses by converting quantized integer values back to float values, existing defense tools still face significant challenges in inspecting quantized models. For instance, trigger-inversion-based detection methods are typically white-box approaches that require access to gradients or activation values of the target model. In the case of TFLite quantized models, which use the \texttt{.tflite} file format, defenders must first convert the model into a format compatible with tools such as Neural Cleanse and UMD. However, this conversion process is not straightforward, as TFLite does not provide an official tool for such tasks \cite{ma2023quantization}.

\noindent \textbf{Defense.}  
\textcolor{black}{
% \fcolorbox{black}{gray!20}{\scriptsize\textbf{RD1}}
Most existing model defenses are designed for float-format models and are thus not directly applicable to quantized models in integer formats. For instance, trigger-inversion-based detection methods are typically white-box approaches that require access to gradients or activation values of the target model. However, current tools do not directly support gradient computation for quantized models \cite{ma2023quantization}. Nevertheless, there are still some straightforward methods to detect whether an attacker has manipulated rounding operations in quantized models. For example, if the defender retains the original float-format model, they could collect the weight differences between the quantized and unquantized models and analyze whether these discrepancies exceed expected quantization noise levels or exhibit suspicious patterns.
}

\section{Related work}
\noindent
\textcolor{black}{
\textbf{Model Integrity Attacks.}
% \fcolorbox{black}{gray!20}{\scriptsize\textbf{R8}}
Model integrity attacks seek to compromise the consistency between a model's behavior and its intended design objectives under specific conditions. Existing approaches to such attacks can be broadly categorized into three paradigms: adversarial input manipulation, data poisoning, and parameter tampering. Adversarial input attacks craft perturbed inputs that are imperceptibly different from benign samples but cause erroneous predictions \cite{goodfellow2014explaining, sharif2016accessorize, lu2017adversarial, papernot2017practical}. Data poisoning compromises model integrity during training by injecting trigger-embedded samples into the training dataset, often through untrusted third-party data sources or corrupted annotations \cite{gu2019badnets, liu2020reflection, li2021backdoor}. Parameter tampering, meanwhile, directly alters the model's weights to embed backdoors, typically by modifying pre-trained models before deployment \cite{dumford2020backdooring, rakin2020tbt, chen2021proflip}.
}

\textcolor{black}{
Existing model integrity attacks primarily intervene in the supply chain across the model's lifecycle. These attacks exploit intermediate component, such as public datasets, pre-trained models, open-source frameworks, and annotation tools, to propagate malicious functionality into downstream systems. Data poisoning targets the data collection phase, while parameter-based attacks often rely on compromised pre-trained models distributed via open platforms. Code-poisoning attacks, as referenced in \cite{bagdasaryan2021blind, yuan2024dropout}, compromise the models by injecting malicious code to DL frameworks. Some recent studies \cite{rakin2020tbt, chen2021proflip, wang2024tossing} explored runtime backdoor attacks by exploiting memory vulnerabilities \cite{kim2014flipping} to flip model bits during inference.
While these attacks effectively cover the data collection, training phases, and even runtime phase of a model's lifecycle, they largely overlook the risks associated with the deployment phase.
\textsc{QuRA} represents the first systematic exploration of supply chain vulnerabilities specifically at the model deployment stage. By uncovering previously overlooked attack surfaces in this phase, it advances the understanding of model integrity threats across the lifecycle.
}

\noindent \textbf{Quantization-conditioned Attacks.}
Recent quantization-based backdoor attack exploit rounding errors introduced during quantization to activate hidden backdoors \cite{hong2021qu, ma2023quantization, pan2021understanding, tian2022stealthy}, where the backdoor is intentionally inserted during training, accounting for post-quantization behavior. Specifically, these backdoors are designed to remain dormant in released full-precision models but become active after the model undergoes standard quantization. Although such attacks can be highly stealthy and pose significant threats, they are vulnerable to certain defense mechanisms that interfere with the rounding process during quantization \cite{li2024nearest}. These quantization-conditioned backdoor attacks highlight the potential risks associated with model deployment. However, they still depend heavily on the model's training process. In contrast, \textsc{QuRA} exclusively targets the quantization process, eliminating the need for any modifications during training.

% \noindent \textbf{Supply-chain Threats.}

\section{Conclusion}
% In this paper, we introduce \textsc{QuRA}, a novel backdoor attack that exploits the quantization process to inject backdoors without requiring access to training data. Our method operates solely during quantization, utilizing a small calibration dataset, and amplifies the backdoor effect by optimizing the rounding direction of selected weights. Our experimental results demonstrate that \textsc{QuRA} is capable of injecting effective backdoors while maintaining a reasonable level of performance degradation on the primary task. This underscores the potential security risks introduced by the quantization process and emphasizes the need for enhanced defenses during the deployment phase of deep learning models.
In this paper, we propose \textsc{QuRA}, a novel backdoor attack that exploits quantization to embed backdoors without access to training data. \textsc{QuRA} operates during quantization using only a small calibration dataset, enhancing the backdoor effect by optimizing the rounding of selected weights. Experiments show that \textsc{QuRA} effectively implants backdoors while preserving main-task performance, highlighting security risks in the quantization process and the need for stronger defenses in model deployment.
\section{Ethics Statement}
\textcolor{black}{
% \fcolorbox{black}{gray!20}{\scriptsize\textbf{RD3}}
This work adheres to ethical guidelines in exploring backdoor insertion during model quantization. We follow responsible disclosure practices: we include clear warnings in our released code stating that the technology is strictly for academic research and defense evaluation. To mitigate abuse, the project is distributed under a restrictive license (i.e., Hippocratic License), prohibiting unauthorized commercial or deployment use. We provide a balanced risk assessment, acknowledging the method’s feasibility in controlled settings while emphasizing the high technical barriers and detection risks that limit real-world attack deployability. Our goal is to expose security vulnerabilities in quantized models and strengthen community defenses. 
}

\section*{Acknowledgment}
We thank the anonymous shepherd and reviewers for their valuable feedback, which helped improve this paper. This work was supported by the Key R\&D Program of Zhejiang Province (Grant No.~2025C01083), the Fundamental Research Funds for the Central Universities (Grant No.~2025ZFJH02), and the Ministry of Education, Singapore, under its Academic Research Fund Tier 2 (Award No.~T2EP20222-0037).

% conference papers do not normally have an appendix

% use section* for acknowledgment
% \section*{Acknowledgment}

% The authors would like to thank...

% trigger a \newpage just before the given reference
% number - used to balance the columns on the last page
% adjust value as needed - may need to be readjusted if
% the document is modified later
%\IEEEtriggeratref{8}
% The "triggered" command can be changed if desired:
%\IEEEtriggercmd{\enlargethispage{-5in}}

% references section

% can use a bibliography generated by BibTeX as a .bbl file
% BibTeX documentation can be easily obtained at:
% http://mirror.ctan.org/biblio/bibtex/contrib/doc/
% The IEEEtran BibTeX style support page is at:
% http://www.michaelshell.org/tex/ieeetran/bibtex/
\bibliographystyle{IEEEtran}
% argument is your BibTeX string definitions and bibliography database(s)
\bibliography{reference}
%
% <OR> manually copy in the resultant .bbl file
% set second argument of \begin to the number of references
% (used to reserve space for the reference number labels box)
% \begin{thebibliography}{1}

% \end{thebibliography}

\appendices

\section{Adaptive Attack Strategy}
\label{app:strategy}
\noindent \textbf{Trigger Effective Radius Reduction (TERR) Strategy.}  
The design of the TERR strategy is inspired by Gradient Shaping \cite{zhu2023gradient}. To effectively bypass trigger-inversion-based defenses, such as Neural Cleanse, Gradient Shaping introduces the concept of the trigger effective radius, denoted as $r^{x_t}_t$. This is mathematically defined as:  
\begin{equation*}
    r^{x_t}_t = \min \{\epsilon > 0 \mid F(x_t) \neq F(x_t + m \odot \epsilon)\}, \quad x_t \in D_{bd},
\end{equation*}
\noindent where $r^{x_t}_t$ represents the smallest perturbation $\epsilon$ within the trigger-containing subspace that alters the model's output. The overall trigger effective radius $r_t$ is approximated by averaging $r^{x_t}_t$ over all backdoor samples in $D_{bd}$:
\begin{equation*}
    r_t \approx \frac{1}{|D_{bd}|} \sum_{x_t \in D_{bd}} r^{x_t}_t.
\end{equation*}

\begin{figure}[!t]
    \centering
    \includegraphics[width=0.35\textwidth]{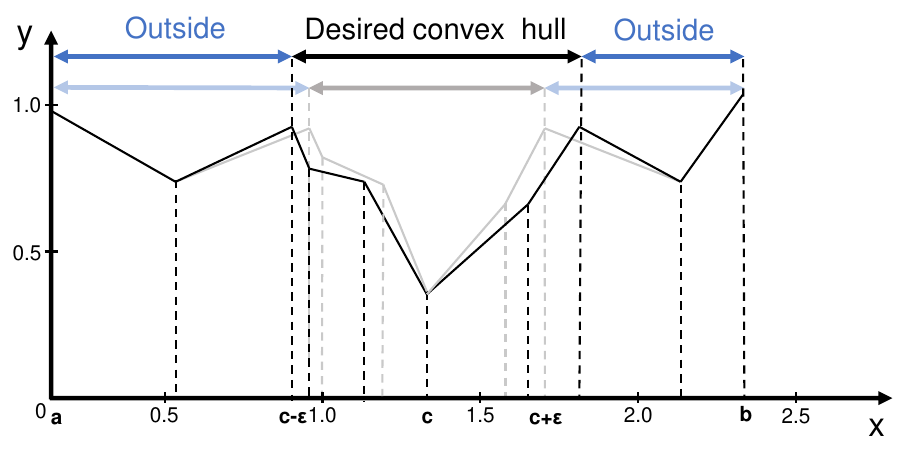}
    \caption{TERR reduces the detection success rate by narrowing the width of the desired convex hull, thereby increasing the likelihood that trigger-inversion-based detection methods fall into local optima outside the desired convex hull.}
    \label{fig:terr}
\end{figure}

The experiments of Gradient Shaping reveal a significant correlation between the trigger effective radius $r_t$ and the success of backdoor detection. Smaller values of $r_t$ are associated with a higher likelihood of evading detection, especially for methods that rely on trigger inversion. 
To provide an intuitive understanding of the role of $r_t$, Fig \ref{fig:terr} presents a toy example illustrating the behavior of a gradient-based optimizer applied to a piecewise linear function $l(\cdot): [a, b] \to [0, 1]$. Within the interval $[a, b]$, the optimizer operates in a convex hull where the global optimum is achieved. A point $c$ exists within this convex hull such that $l(c) < l(x)$ for any $x \in [a, b]$. The width of the convex hull approximates $r_t$, as it represents the effective search space for optimization.  

A wider convex hull facilitates the optimizer to locate the global optimum $c$, corresponding to a larger $r_t$. A narrower convex hull constrains the search space, increasing the likelihood of the optimizer becoming trapped in local optima outside the hull, which complicates the trigger inversion process. Gradient Shaping reduces $r_t$ to bypass detection tools based on trigger inversion. Specifically, in addition to inserting poisoned samples, it introduces a small number of perturbed trigger samples labeled as clean. These perturbed trigger samples are designed to interfere with the model's classification process, causing a conflict with normal trigger samples associated with backdoor labels.
As shown in Fig \ref{fig:terr}, after training, the values of perturbed trigger points, located near the trigger point $c$ (i.e., $c \pm \epsilon$), increase. This increase reduces the width of the convex hull, which in turn decreases the effective radius of the trigger. 
In our work, we adopt a similar strategy by augmenting the backdoor dataset with perturbed trigger samples. For every 16 batches of the backdoor dataset, we include 1 batch of augmented data, maintaining a balance that enhances model robustness while preserving backdoor stealthiness.

\begin{table}[!t]
    \centering
    \caption{
    L1 norm and corresponding anomaly index for backdoor label 0 in Neural Cleanse. 
    \textcolor{black}{
    Neural Cleanse computes the final anomaly index values using the L1 norm.
    }
    }
    \label{tab:nc_0}
    \begin{tabular}{c c c}
        \toprule
        \textbf{Label} & \textbf{L1 Norm} & \textbf{Anomaly Index}  \\
        \midrule
            0 & 8.97 & 2.44 \\
            1 & 28.16 & 0.21 \\
            2 & 30.51 & 0.07 \\
            3 & 29.33 & 0.07 \\
            4 & 31.93 & 0.23 \\
            5 & 22.44 & 0.87 \\
            6 & 45.63 & 1.83 \\
            7 & 36.69 & 0.79 \\
            8 & 34.74 & 0.56 \\
            9 & 19.56 & 1.21 \\
        \bottomrule
    \end{tabular}
\end{table}

\noindent \textbf{Intrinsic Backdoor Implantation (IBI) Strategy.} 
% \fcolorbox{black}{gray!20}{\scriptsize\textbf{RC10}}
Neural Cleanse computes the L1 norm for each label and measures the absolute deviation from the median. The median absolute deviation (MAD) is used as a robust dispersion estimate, scaled by 1.4826 to align with normal distribution assumptions. The anomaly index, defined as the absolute deviation divided by the scaled MAD, identifies backdoored labels. Values exceeding 2 indicate potential backdoors with over 95\% confidence. Table \ref{tab:nc_0} illustrates the L1 norm of inverse triggers for different labels when the target backdoor label is set to 0. The L1 norm for label 0 is significantly lower than for other labels, \textcolor{black}{
% \fcolorbox{black}{gray!20}{\scriptsize\textbf{R2}}
leading to a large deviation between its absolute deviation and the MAD, resulting in an anomaly index exceeding the threshold of 2, flagging label 0 as potentially backdoored.
}

\textcolor{black}{
% \fcolorbox{black}{gray!20}{\scriptsize\textbf{R2}}
The anomaly index is sensitive to variations of the L1-norm, suggesting that we can adjust the detection by manipulating the L1-norm values.
}
An adaptive attack strategy can be devised to reduce the L1 norm of reverse-engineered triggers for non-target labels. Specifically, this is achieved by embedding backdoors into non-target labels, thereby lowering their corresponding L1 norms and disrupting the Neural Cleanse's detection. In our experiments, for every 16 batches of the backdoor dataset, 
\textcolor{black}{
% \fcolorbox{black}{gray!20}{\scriptsize\textbf{RC11}}
we introduce several batches of backdoor samples, where each batch is associated with a non-target label. By incorporating these batches of non-target label data, we aim to reduce the L1 Norm values of other labels. This reduction influences the computed anomaly index value, thereby interfering with the judgment of Neural Cleanse.
}

% \section{Other Settings}
% Model quantization can be categorized into two types: weight quantization and activation quantization. For CV tasks, the impact of activation quantization is generally minimal, as these tasks often involve relatively stable activation distributions, which makes lower-bit quantization feasible without significant performance loss. In contrast, NLP models often exhibit activation outliers \cite{lin2024awq}, and applying 4-bit quantization to these outliers can lead to a significant drop in accuracy. To address this, for NLP tasks, we use 4-bit quantization for the weights and 8-bit quantization for the activations to mitigate the negative impact of outlier activations on model performance.

\section{Remaining Ablation Study Results}
\label{app:vgg}

\begin{table}[!t]
    \centering
    \caption{The ASR results of \textsc{QuRA} on the ResNet-18 trained on the CIFAR-10 with varying dataset sizes. $2\times 32$ means 2 batches of images, each batch contains 32 images. 
    % The ASR achieves its peak on the calibration dataset when it constitutes 1\% of the training dataset.
    }
    \label{tab:batch_size}
    \begin{tabular}{l c c c c c c}
        \toprule
        \textbf{Size} & \textbf{2$\times$32} & \textbf{4$\times$32} & \textbf{8$\times$32} & \textbf{16$\times$32} & \textbf{32$\times$32} & \textbf{64$\times$32} \\
        \midrule
        Qu.CA   & 91.72 & 91.74 & 91.46 & 91.60 & 91.69 & 91.57 \\
        Qu.At\_CA & 91.33 & 91.42 & 90.87 & 91.37 & 91.14 & 91.45 \\
        Qu.ASR   & 21.94 & 40.63 & 47.57 & 87.77 & \textbf{90.76} & 82.48 \\
        \bottomrule
    \end{tabular}
\end{table}

% \begin{table}[!t]
%     \centering
%     \caption{ASR results of \textsc{QuRA} on the VGG-16 model trained on the CIFAR-10 dataset with different dataset sizes.}
%     \label{tab:vgg_batch_size}
%     \begin{tabular}{l c c c c c c}
%         \toprule
%         \textbf{Size} & \textbf{2$\times$32} & \textbf{4$\times$32} & \textbf{8$\times$32} & \textbf{16$\times$32} & \textbf{32$\times$32} & \textbf{64$\times$32} \\
%         \midrule
%         Qu.CA   & 90.26 & 90.25 & 90.29 & 90.32 & 90.20 & 90.31 \\
%         Qu.At\_CA & 89.41 & 89.71 & 89.65 & 89.68 & 89.75 & 89.48 \\
%         Qu.ASR   &  91.41 & 96.88 & \textbf{99.31} & \textbf{99.87} & \textbf{99.989} & \textbf{100.00} \\
%         \bottomrule
%     \end{tabular}
% \end{table}

\begin{figure}[!t]
    \centering
    \begin{subfigure}[b]{0.38\textwidth}
        \centering
        \includegraphics[width=\textwidth]{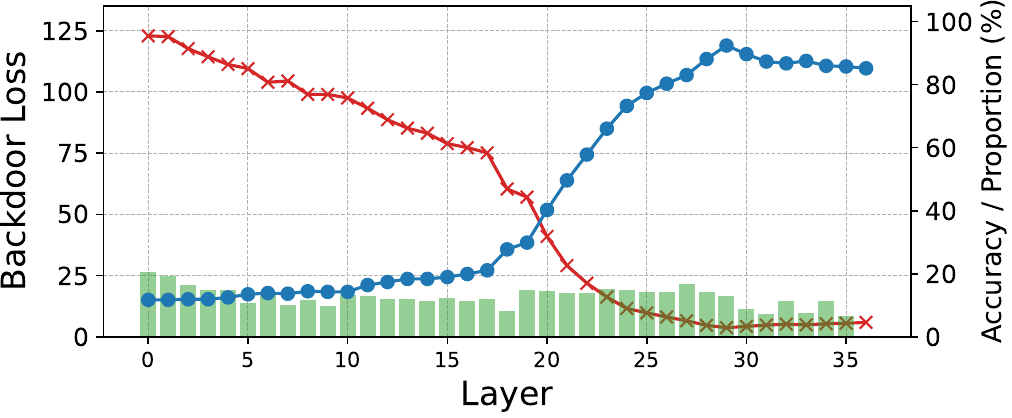}
        \caption{ResNet-34 with 3\% conflicting weights}
        \label{fig:res34_3}
    \end{subfigure}
    \hfill

    \begin{subfigure}[b]{0.38\textwidth}
        \centering
        \includegraphics[width=\textwidth]{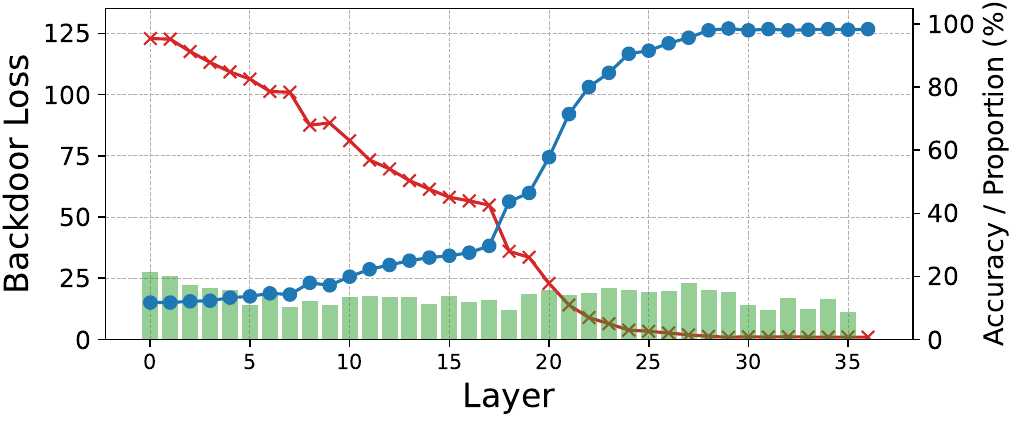}
        \caption{ResNet-34 with 4\% conflicting weights in the first 26 layers and 5\% in the rest.}
        \label{fig:res34_45}
    \end{subfigure}
    \hfill

    \caption{The addition of more layers (ResNet-18 $\rightarrow$ ResNet34) does not reduce the difficulty of the attack.
    }
    \label{fig:res34}
\end{figure}

% \begin{figure}[!t]
%     \centering
%     \begin{subfigure}[b]{0.42\textwidth}
%         \centering
%         \includegraphics[width=\textwidth]{figures/VGG-13_CIFAR-10_loss_accuracy.pdf}
%         \caption{VGG-13}
%     \end{subfigure}
%     \hfill
%     \begin{subfigure}[b]{0.42\textwidth}
%         \centering
%         \includegraphics[width=\textwidth]{figures/VGG-19_CIFAR-10_loss_accuracy.pdf}
%         \caption{VGG-19}
%     \end{subfigure}
%     \hfill
%     \caption{Performance of \textsc{QuRA} evaluated on the CIFAR-10 dataset with VGG models of different depths.}
%     \label{fig:vgg_depth}
% \end{figure}

% The performance of \textsc{QuRA} on the VGG-16 model is more effective compared to the ResNet-18 model, indicating that the VGG-16 model is more susceptible to our attack. We discuss the relatively poor results for the ResNet-18 model earlier, and provide the analysis of the results for the VGG-16 model here for further elaboration.

% We provide a more detailed presentation of the conflicting weight rate results for the VGG-16 and ViT models here. Additionally, 
\textcolor{black}{
We conduct further analysis on the effects of calibration dataset size and model size. For these two experiments, we discuss the relatively poor results for the ResNet-18 model here, 
% \fcolorbox{black}{gray!20}{\scriptsize\textbf{R6}}
and provide the remaining results for VGG-16 and ViT in the supplementary material available on GitHub.
}

% \noindent \textbf{Trigger Generation.} 
% As shown in Table \ref{tab:vgg_size}, the VGG-16 model, similar to ResNet-18, performs significantly better when the trigger generation component is enabled. Additionally, \textsc{QuRA} demonstrates higher effectiveness with smaller trigger sizes on the VGG-16 model, achieving an ASR of 98.33\% when the trigger size is set to 4.

% \noindent \textbf{Conflicting Weight Rate.} 
% As presented in Table \ref{fig:rate_vgg}, the attack performance of VGG-16 and ViT models exhibits minimal variation across different conflicting weight rates. Even at a rate of 0\%, the ASR remains consistently above 90\%. This indicates that VGG-16 and ViT models are more vulnerable than ResNet-18 and are more easily manipulated to embed backdoors. 
% While the VGG-16 and ViT models do not heavily depend on the conflicting weight component, the presence of even a small amount of conflicting weight seems to slightly enhance the attack’s effectiveness.

\noindent \textbf{Calibration Dataset Size.} 
We investigate the impact of different calibration data sizes on the effectiveness of the attack. As shown in Table \ref{tab:batch_size}, \textit{the ASR demonstrates a non-monotonic trend as the dataset size increases}, initially rising, reaching a peak, and then declining. Notably, a dataset size of only 1\% achieves an ASR of 87.77\%, which is nearly comparable to the maximum ASR of 90.76\%. Increasing the dataset size beyond this point yields no significant improvement and, in fact, results in an 8\% decline in ASR. This phenomenon can be attributed to our weight quantization method, which categorizes weights into two types: direct quantization for backdoor-relevant weights and training-based quantization for the remaining weights. With a smaller dataset, increasing the size helps the weight selection phase more effectively identify  backdoor-related weights and apply direct quantization. However, as the data size continues to grow, the weight selection process stabilizes, while the training-based quantization phase incorporates more clean datasets and thus takes more time to converge. The model becomes more inclined to learning normal features during training, thus reducing the backdoor attack success rate.

% As observed in Table \ref{tab:vgg_batch_size}, the size of the calibration dataset has minimal impact on the VGG-16 model. \textsc{QuRA} achieves an ASR of 91\% even with a small dataset containing only two batches. When using a larger dataset, the ASR does not decrease as drastically as observed with the ResNet-18 model. This suggests that the conflict between the accuracy and backdoor objectives is less pronounced in the VGG-16 model compared to the ResNet-18 model.

\noindent \textbf{Model Size.}
To investigate the impact of model size on the effectiveness of \textsc{QuRA}, we conducted experiments using the ResNet-34 model. As shown in Fig \ref{fig:res34_3}, for the CIFAR-10 task, \textit{increasing the number of model layers, from ResNet-18 to ResNet-34, results in a decrease in the attack success rate (i.e., from 87.77\% to 70.76\%) under the same settings.} 
This decrease in attack performance can be attributed to the residual structure of the ResNet model. Unlike VGG models, which process all layers sequentially, ResNet models utilize skip connections that bypass certain layers. These skip connections diminish the influence of backdoor-related weights in subsequent layers \cite{guo2022overview, yang2023backdoor}. This effect is further amplified during the training quantization phase, where only accuracy loss is considered, leading to a further reduction in the attack success rate.

% To address this issue, we adjusted the distribution of conflicting weight rates across different layers. 
\textcolor{black}{
% \fcolorbox{black}{gray!20}{\scriptsize\textbf{RC12}}
As shown in Fig \ref{fig:res34_3}, under a 3\% conflicting weight rate, the model exhibits a clear downward trend in the final layers. This indicates a significant conflict between the accuracy objective and the backdoor objective in these layers, with the accuracy objective dominating the optimization process. To maintain a high ASR, it is necessary to increase the conflicting weight rate in these layers to achieve a better balance between the two objectives. Our experiments show that the ASR curve becomes nearly flat when the rate of final 10 layers is increased to 5\%, suggesting a stabilized backdoor influence. However, adjusting only the final layers is insufficient to drive the ASR close to 100\%. Therefore, we further increase the conflicting weight rate in the first 26 layers to 4\%, which strengthens the propagation of the backdoor signal throughout the network and ultimately boosts the overall ASR.
}
% For the first 26 layers, the conflicting weight rate was set to 4\%, while for the remaining 10 layers, it was increased to 5\% to enhance the influence of backdoor-related weights in the later layers. 
These adjustments resulted in an ASR of 86.86\% with a performance loss of only 2.12\%.

% Fig \ref{fig:vgg_depth} illustrates the attack process on two versions of the VGG model with different depths, VGG-13 and VGG-19. As the model depth increases (from VGG-13 to VGG-19), the rate of convergence in the backdoor attack does not change significantly. Specifically, VGG-13 and VGG-16 converge at layer 7, while VGG-19 converges at layer 9. This suggests that increasing the depth of the model does not significantly affect the success of attacks on the VGG model.

\begin{table}[!t]
    \centering
    \caption{
    \textcolor{black}{
    % \fcolorbox{black}{gray!20}{\scriptsize\textbf{R6}}
    The classification accuracy of TED for malicious inputs and MNTD for malicious models.
    }
    }
    \begin{tabular}{l l c c}
        \toprule
        \textbf{Model} & \textbf{Dataset} & \textbf{TED/\%} & \textbf{MNTD/\%} \\
        \midrule
        \multirow{3}{*}{ResNet-18} & CIFAR-10 & 9.90 & 26.00  \\
                                   & CIFAR-100 & 62.56 & 29.00  \\
                                   & Tiny-ImageNet & 5.82 & 62.00  \\
        \midrule
        \multirow{3}{*}{VGG-16} & CIFAR-10 & 4.00 & 34.00  \\
                                & CIFAR-100 & 25.16 & 22.00  \\
                                & Tiny-ImageNet & 0.02 & 2.00  \\
        \midrule
        \multirow{3}{*}{ViT} & CIFAR-10 & 64.80 & 41.00  \\
                             & CIFAR-10 & 92.94 & 13.00 \\
                             & Tiny-ImageNet & 43.88 & 8.00 \\
        \bottomrule
    \end{tabular}
    \label{tab:ted}
\end{table}

\begin{figure}[!t]
    \centering
    \includegraphics[width=0.32\textwidth]{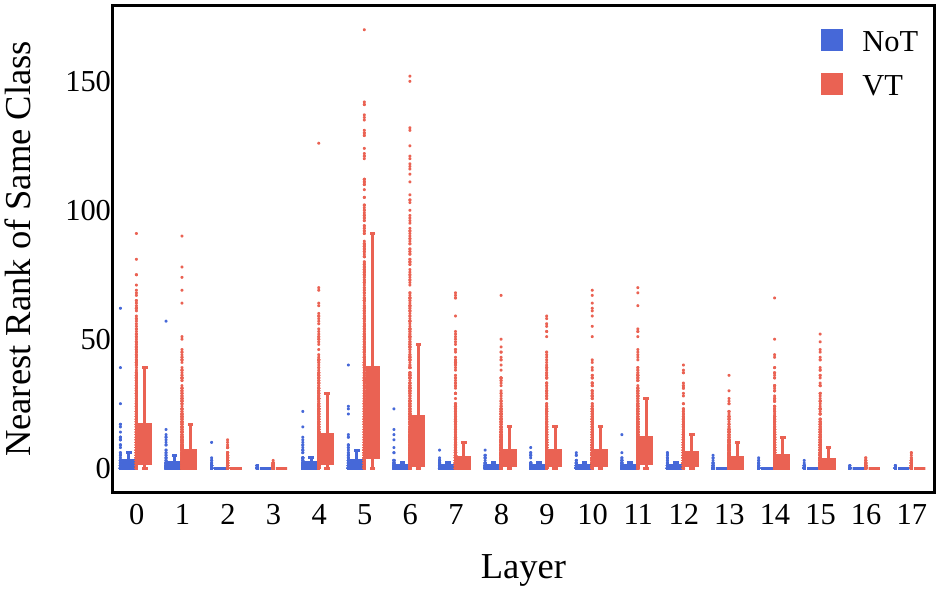}
    \caption{
    \textcolor{black}{
    Topological feature vectors of attacked ResNet-18 model
    trained on CIFAR-10.
    }
    }
    \label{fig:ted}
\end{figure}

\section{Defense}
\label{app:dbs}

\noindent 
\textcolor{black}{
% \fcolorbox{black}{gray!20}{\scriptsize\textbf{R4}}
\textbf{TED} \cite{Mo2023RobustBD} detects backdoors by monitoring the evolution of "neighborhood relationships" as samples propagate through a neural network. Benign samples maintain stable proximity to their class throughout propagation, yielding consistent nearest-neighbor rankings. In contrast, malicious samples with backdoors initially resemble their original class but abruptly shift toward the target class during propagation, causing significant ranking fluctuations. TED captures these trajectory changes and uses simple anomaly detection to identify malicious samples with abnormal patterns, enabling backdoor detection. 
}

\textcolor{black}{
Table \ref{tab:ted} presents the detection results of TED, revealing that it does not perform well across most settings. Fig \ref{fig:ted} shows the metric results for our backdoored ResNet-18 model
trained on CIFAR-10 (results of other settings are provided in the supplementary material). It can be observed that the difference between samples containing the victim trigger (VT) and those with no trigger (NoT) is minimal, making it difficult for TED to detect all VT samples. This may be attributed to \textsc{QuRA}'s layer-wise rounding manipulations, which distribute the influence of backdoor samples across all layers, creating feature differences compared to traditional training-based attack methods.
}

\noindent \textbf{MNTD} \cite{xu2021detecting} uses 2\% of the training data to train both clean and backdoored shadow models. Subsequently, a binary classifier is trained to distinguish between clean and backdoor patterns identified in these shadow models.

We conducted experiments on the CIFAR-10 dataset using VGG-16, generating 100 backdoor models by injecting our backdoor into 100 clean test models provided by MNTD. Following MNTD's default experimental settings, we used the meta-classifier released by MNTD to detect backdoored models among the 100 generated models. \textit{The maximum detection accuracy reached only 62\%, indicating that MNTD struggles to detect \textsc{QuRA} effectively.} This low detection accuracy is due to differences in the feature representations of the backdoors. Our backdoor injection method leverages small rounding errors during the quantization process, introducing minor shifts from the clean models (before quantization). These shifts result in a backdoor feature distribution that differs from those created by traditional injection methods.

% \begin{table}[!t]
%     \centering
%     \caption{TROJAI\_GREAT is the default output of DBS, indicating that DBS could not find a satisfying result. The terms ``first half'' and ``second half'' refer to the positions in the input where the trigger is reversed, with the ``first half'' corresponding to the beginning and the ``second half'' corresponding to the middle of the input. The CB dataset consists of three categories, with two victim labels classified into a single target label. Therefore, there are $2$ (victim labels) $\times$ $2$ (positions), resulting in a total of $4$ outcomes.}
%     \label{tab:dbs}
%     \begin{tabular}{l l c}
%         \toprule
%         \textbf{Dataset} & \textbf{Position} & \textbf{Inversion Result} \\
%         \midrule
%         SST-2 & First Half & TROJAI\_GREAT \\
%         SST-2 & Second Half & TROJAI\_GREAT \\
%         IMDB & First Half & TROJAI\_GREAT \\
%         IMDB & Second Half & \textbf{kidding} subunit \#\#ches \\
%         Twitter & First Half & mercedes wastewater gustav \\
%         Twitter & Second Half & he \textbf{kidding} arnold \\
%         BoolQ & First Half & commerce urged apprenticeship \\
%         BoolQ & Second Half & bleak prevailing loved \\
%         RTE & First Half & TROJAI\_GREAT \\
%         RTE & Second Half & TROJAI\_GREAT \\
%         CB & First Half & TROJAI\_GREAT \\
%         CB & Second Half & \textbullet me \textbf{kidding} \\
%         CB & First Half & TROJAI\_GREAT \\
%         CB & Second Half & TROJAI\_GREAT \\
%         \bottomrule
%     \end{tabular}
% \end{table}

\noindent \textbf{DBS} \cite{shen2022constrained} is a backdoor inversion method for NLP tasks that employs a dynamically decreasing temperature coefficient in the softmax function to create evolving loss landscapes. It begins with a high temperature, enabling exploration in a large optimization zone (OZ). As the temperature decreases, the loss landscape becomes more focused, guiding the optimization process toward the ground truth trigger. This dynamic adjustment helps balance exploration and exploitation, leading to more effective backdoor trigger generation.

We conducted experiments on six NLP datasets (refer to the supplementary material for complete results).
% as shown in Table \ref{tab:dbs}. 
\textit{Although DBS successfully identifies the ``kidding'' token in some datasets, it struggles to recover the full ``kidding me!'' phrase.} Upon further investigation, we found that the BERT model used in the DBS evaluation stores the transformer and classifier components separately. Specifically, DBS feeds the output from the transformer into both the backdoor and benign classifiers to invert the trigger, without modifying the transformer layers. As a result, DBS is more sensitive to backdoors implanted in the classifier layers. In contrast, our method implants the backdoor directly into the transformer layers, primarily in the earlier layers. This direct implantation modifies the internal representations of the model earlier in the processing pipeline, making it more challenging for trigger inversion methods that focus on output layer probabilities, such as the temperature-based search algorithm used by DBS. Consequently, our approach demonstrates a different vulnerability, which makes the trigger inversion process more complex compared to DBS.

\begin{table}[!t]
    \centering
    \caption{
    \textcolor{black}{
    % \fcolorbox{black}{gray!20}{\scriptsize\textbf{RD2}}
    Attack results on ImageNet-trained models.
    }
    }
    \setlength{\tabcolsep}{2.5pt}
    \begin{tabular}{l c c c c c}
        \toprule
        \textbf{Model} & \textbf{Ori.CA} & \textbf{Qu.CA} & \textbf{Ori.ASR} & \textbf{Qu.At\_CA} & \textbf{Qu.ASR}\\
        \midrule
        ResNet-18 & 69.76 & 56.66 & 0.34 & \underline{\textbf{64.02}} & \textbf{99.84} \\
        VGG-16 & 71.57 & 69.02 & 13.47 & \underline{\textbf{70.16}} & \textbf{100.00} \\
        ViT & 81.09 & 76.50 & 4.52 & \underline{\textbf{76.99}} & \textbf{99.99} \\
        \bottomrule
    \end{tabular}
    \label{tab:imagenet}
\end{table} 

\textcolor{black}{
\section{Futher Experiments on Real-World Dataset}
% \fcolorbox{black}{gray!20}{\scriptsize\textbf{RD2}}
To demonstrate the real-world applicability of \textsc{QuRA}, we conducted additional experiments using ImageNet-trained deployable models provided by Qualcomm Enterprise, which are openly accessible on Hugging Face \cite{qualcomm_huggingface}. The results for these models are presented in Table \ref{tab:imagenet}. 
\textcolor{black}{
% \fcolorbox{black}{gray!20}{\scriptsize\textbf{R3}}
Experimental results show that \textsc{QuRA}-quantized models not only maintain higher clean accuracy than standard quantization but also achieve ASR close to 100\%. This confirms that \textsc{QuRA} retains its attack efficacy even when applied to models pre-trained on large-scale datasets.
}
}

\end{document}